\documentclass[journal,compsoc,twoside,lettersize]{IEEEtran}

\ifCLASSOPTIONcompsoc
  \usepackage[nocompress]{cite}
\else
  \usepackage{cite}
\fi

\usepackage{amsmath}
\usepackage{mathrsfs}
\usepackage{diagbox,xspace}

\makeatletter		

\newcommand{\Rmnum}[1]{\expandafter\@slowromancap\romannumeral #1@}
\makeatother

\usepackage{algorithm}
\usepackage{algorithmic}
\usepackage{booktabs}
\usepackage{makecell}
\usepackage{amssymb}
\usepackage{bm}		
\usepackage[table]{xcolor}
\usepackage[colorlinks, citecolor=blue]{hyperref} 
\usepackage{threeparttable}

\usepackage{colortbl}
\usepackage{nicematrix}
\usepackage{tabularx}
\usepackage{multirow}
\usepackage{dsfont}
\usepackage{enumitem}
\usepackage{eucal}
\usepackage[american]{babel}
\usepackage{ragged2e}

\newcolumntype{L}[1]{>{\raggedright\arraybackslash}p{#1}}
\newcolumntype{C}[1]{>{\centering\arraybackslash}p{#1}}
\newcolumntype{R}[1]{>{\raggedleft\arraybackslash}p{#1}}

\usepackage{amsthm}
\newtheorem{definition}{Definition}

\usepackage{subfigure}

\usepackage{textcomp}
\usepackage{stfloats}
\usepackage{url}
\usepackage{verbatim}
\usepackage{graphicx}
\def\BibTeX{{\rm B\kern-.05em{\sc i\kern-.025em b}\kern-.08em
    T\kern-.1667em\lower.7ex\hbox{E}\kern-.125emX}}
\usepackage{balance}

\usepackage{algorithm}  
\usepackage{fontawesome}

\usepackage{makecell}
\usepackage{mathtools}

\usepackage{booktabs}
\usepackage{xspace}
\usepackage{colortbl}
\usepackage{xcolor}
\usepackage{multirow}
\usepackage{tablefootnote}
\usepackage{url}

\usepackage{bm}
\usepackage{enumitem}
\usepackage{bbding}

\usepackage{array}         
\usepackage{booktabs}      
\usepackage{makecell}      
\usepackage{tabularx}      
\usepackage{caption}       

\usepackage{subcaption}
\usepackage{times}
\usepackage{amsmath}
\usepackage{amssymb}

\usepackage{lipsum}
\usepackage{float}

\usepackage{cite}
\usepackage{amsmath,amssymb,amsfonts}
\usepackage{textcomp}
\usepackage{adjustbox}
\usepackage{tablefootnote}

\newcommand{\upgood}{\color{green!70!black}$\Uparrow$}
\newcommand{\downgood}{\color{red!70!black}$\Downarrow$}

\newcolumntype{o}{>{\columncolor{red!5}}c}
\newcolumntype{b}{>{\columncolor{blue!5}}c}
\newcolumntype{g}{>{\columncolor{lightgray!30}}c}

\newcolumntype{i}{>{\columncolor{red!15}}c}
\newcolumntype{n}{>{\columncolor{blue!15}}c}

\usepackage{multirow, hhline}

\newcommand{\myparagraph}[1]{\vspace{1ex}\noindent\textbf{#1.}\hspace{1em}}

\newcommand{\runhao}{\textcolor{black}{}}

\newcommand{\lightea}{\textsf{LightTEA}\xspace}
\newcommand{\dualmatch}{\textsf{DualMatch}\xspace}

\newcommand{\mraea}{\textsf{MRAEA}\xspace}
\newcommand{\rrea}{\textsf{RREA}\xspace}

\newcommand{\mgtea}{\textsf{MGTEA}\xspace}

\newcommand{\yw}{\texttt{YAGO-WIKI50K}\xspace}
\newcommand{\ywa}{\texttt{YAGO-WIKI50K-1K}\xspace}
\newcommand{\ywb}{\texttt{YAGO-WIKI50K-5K}\xspace}
\newcommand{\ywc}{\texttt{YAGO-WIKI20K}\xspace}
\newcommand{\ywas}{\texttt{YW1K}\xspace}
\newcommand{\ywbs}{\texttt{YW5K}\xspace}

\newcommand{\icewswiki}{\texttt{ICEWS-WIKI}\xspace}
\newcommand{\icewsyago}{\texttt{ICEWS-YAGO}\xspace}

\newcommand{\MTransE}{\textsf{MTransE}\xspace}
\newcommand{\AlignE}{\textsf{AlignE}\xspace}
\newcommand{\BootEA}{\textsf{BootEA}\xspace}
\newcommand{\GCNAlign}{\textsf{GCN-Align}\xspace}
\newcommand{\RDGCN}{\textsf{RDGCN}\xspace}
\newcommand{\DualAMN}{\textsf{Dual-AMN}\xspace}
\newcommand{\Dualmatch}{\textsf{Dual-Match}\xspace}
\newcommand{\TEAGNN}{\textsf{TEA-GNN}\xspace}
\newcommand{\TREA}{\textsf{TREA}\xspace}
\newcommand{\TEA}{\textsf{TEA}\xspace}
\newcommand{\STEA}{\textsf{STEA}\xspace}
\newcommand{\BERT}{\textsf{BERT}\xspace}
\newcommand{\FuAlign}{\textsf{FuAlign}\xspace}
\newcommand{\BERTINT}{\textsf{BERT-INT}\xspace}
\newcommand{\PARIS}{\textsf{PARIS}\xspace}
\newcommand{\SimpleHHEA}{\textsf{Simple-HHEA}\xspace}
\newcommand{\ChatEA}{\textsf{ChatEA}\xspace}

\newcommand{\MGTEA}{\textsf{MGTEA}\xspace}
\newcommand{\HTEA}{\textsf{HTEA}\xspace}

\newcommand{\nativerag}{\textsf{Naive RAG}\xspace}
\newcommand{\selfrag}{\textsf{Self-RAG}\xspace}
\newcommand{\self}{\textsf{Self-Consistency}\xspace}

\newcommand{\icews}{\texttt{ICEWS}\xspace}
\newcommand{\icewsa}{\texttt{ICEWS05-15}\xspace}
\newcommand{\yago}{\texttt{YAGO}\xspace}
\newcommand{\wikidata}{\texttt{Wikidata}\xspace}
\newcommand{\dicews}{\texttt{DICEWS}\xspace}
\newcommand{\da}{\texttt{DICEWS-200}\xspace}
\newcommand{\db}{\texttt{DICEWS-1K}\xspace}
\newcommand{\das}{\texttt{D200}\xspace}
\newcommand{\dbs}{\texttt{D1K}\xspace}
\newcommand{\enfr}{\texttt{DBP15K(EN-FR)}\xspace}
\newcommand{\dbpwiki}{\texttt{DBP-WIKI}\xspace}

\definecolor{c1}{HTML}{344C11}
\definecolor{c2}{HTML}{7e0f12}

\newcommand{\wild}{\texttt{WildBETA}\xspace}
\newcommand{\Beta}{\texttt{BETA}\xspace}
\newcommand{\ourmodel}{\textsf{HyDRA}\xspace}

\usepackage{amssymb,utfsym}

\newcommand{\cmark}{{\textcolor{c1}\checkmark}}
\newcommand{\tmark}{\mathrel{\cmark\textcolor{c2}{\mkern-13mu \raisebox{0.45ex}{$\smallsetminus$}}}} 

\newcommand{\xmark}{{\textcolor{c2}\XSolidBrush}}
\newcommand{\squishlist}{
 \begin{list}{$\bullet$}
 { \setlength{\itemsep}{0pt}
   \setlength{\parsep}{3pt}
   \setlength{\topsep}{3pt}
   \setlength{\partopsep}{0pt}
   \setlength{\leftmargin}{1.2em}
   \setlength{\labelwidth}{1em}
   \setlength{\labelsep}{0.6em}
 }
}
\newcommand{\squishend}{
 \end{list}
}

\begin{document}
%
\title{
Towards Temporal Knowledge Graph Alignment in the Wild}

\author{Runhao Zhao, Weixin Zeng, Wentao Zhang, Xiang Zhao, Jiuyang Tang, and Lei Chen,~\IEEEmembership{Fellow,~IEEE}

\IEEEcompsocitemizethanks{\IEEEcompsocthanksitem Manuscript received June 30, 2025.  
	\IEEEcompsocthanksitem Runhao Zhao, Weixin Zeng, Xiang Zhao, Jiuyang Tang are with Laboratory for Big Data and Decision, National University of Defense Technology, China. (e-mail: \{runhaozhao, zengweixin13, xiangzhao, jiuyang\_tang\}@nudt.edu.cn).
	\IEEEcompsocthanksitem 
	Wentao Zhang is with the Center for machine learning research, Peking University, Beijing 100871, China (e-mail: wentao.zhang@pku.edu.cn).
    \IEEEcompsocthanksitem Lei Chen is with the Hong Kong University of Science and Technology, Clear Water Bay, Hong Kong (e-mail: leichen@cse.ust.hk).
}

\thanks{ 
Corresponding author: Wentao Zhang and Xiang Zhao.
}

\thanks{ 
Runhao Zhao and Weixin Zeng contributed equally to this work.
}

}

\markboth{Manuscript Submitted to Journal; June~2025}%
{Shell \MakeLowercase{\textit{Zhao et al.}}: Temporal Knowledge Graph Alignment in the Wild}

\IEEEtitleabstractindextext{%
\begin{abstract}
\justifying
Temporal Knowledge Graph Alignment (TKGA) seeks to identify equivalent entities across heterogeneous temporal knowledge graphs (TKGs) for fusion to improve their completeness.
Although some approaches have been proposed to tackle this task, most assume unified temporal element standards and simplified temporal structures across different TKGs. They cannot deal with TKGA in the wild (TKGA-Wild), where \emph{multi-scale temporal element entanglement (i.e., multi-granular temporal coexistence and temporal interval topological disparity)} and \emph{cross-source temporal structural imbalances (i.e., multi-source temporal incompleteness and temporal event density imbalance)} are common. To bridge this gap, we study the task of TKGA-Wild and propose \ourmodel, a new and effective solution. \ourmodel is the first to reformulate the task via \emph{multi-scale hypergraph retrieval-augmented generation} to address the challenges of TKGA-Wild. \ourmodel effectively captures complex structural dependencies, models multi-granular temporal features, and mitigates temporal disparities.
In addition, we design a new \emph{scale-weave synergy} mechanism for \ourmodel, which incorporates \emph{intra-scale interactions} and cross-scale \emph{conflict detection}. This mechanism is designed to alleviate the fragmentation caused by \emph{multi-source temporal incompleteness} and resolves \emph{inconsistencies arising from complex and uneven temporal event density distributions}, thereby enhancing the model's capacity to handle the intricacies of real-world temporal alignment.
Finally, there is no standard benchmark that captures these challenges of TKGA-Wild and effectively evaluates existing methods. To this end, we formally propose to benchmark challenges for TKGA-Wild and validate the effectiveness of the method by establishing two new datasets, i.e., \Beta and \wild. Extensive experiments on the new datasets and six representative benchmarks show that \Beta and \wild better reflect real-world challenges. Meanwhile, \ourmodel proposes a new paradigm for TKGA-Wild, consistently outperforming 24 competitive baselines, achieving up to a 43.3\% improvement in Hits@1, while maintaining strong efficiency and scalability.
\end{abstract}

}

\maketitle

\IEEEdisplaynontitleabstractindextext

%
\IEEEpeerreviewmaketitle

\section{Introduction}
\label{sec:introduction}
\IEEEPARstart{A}{s} real-world facts evolve over time, static knowledge graphs (KGs) are inadequate for time-sensitive tasks like question answering~\cite{RetrievalQA} and knowledge reasoning~\cite{TPAMI6,ICLRreasoning25}. 
Thus, temporal knowledge graphs (TKGs) have gained increasing attention.
A fact in TKG in expressed in the form of a quadruple $(e_h,r,e_t,\tau)$, where $\tau$ is the time interval with a beginning timestamp $t_b$ and an ending timestamp $t_e$. 
If $\tau$ is not none, the fact is referred to as a \emph{valid quadruple} or a \emph{temporal fact}.

To increase the coverage and also eliminate the outdated facts in TKGs, a critical step is temporal knowledge graph alignment (TKGA), which aims to match TKGs constructed from different sources and merge relevant facts using aligned entities as anchors~\cite{wsdm,TREA}. 
The main focus in TKGA is the alignment of entities (TEA), as the number of entities (usually over 10k) is much larger than that of relations (usually in hundreds)~\cite{TREA}. An example of TEA is illustrated in Fig.~\ref{introduction2}, which aims to match \texttt{Luis Milla} in $TKG_1$ and \texttt{<Luis\_Milla>} in $TKG_2$.

To address TKGA, several methods have been proposed~\cite{TEA-GNN,lightea,dualmatch,simplehhea,wsdm}.
While some aim to inject temporal information into the entity representations~\cite{TREA,Tem-EA,TGA-EA}, others directly use the time information for alignment~\cite{lightea,STEA,dualmatch}. These methods often assume \textbf{unified temporal element standards} and \textbf{simplified temporal structures} (e.g., Fig.~\ref{introduction2}).
However, these assumptions are often unrealistic. In fact, real-world data like \icewsa\footnote{\url{https://www.andybeger.com/icews}}, \wikidata\footnote{\url{https://www.wikidata.org/}}, and \yago\footnote{\url{https://yago-knowledge.org/}} record multi-granular temporal information, ranging from macro-level intervals (e.g., centuries) to fine-grained timestamps (e.g., specific dates). \icewsa~\cite{nip24}, focusing on political crisis events, covers 2005-2015, while \yago~\cite{YAGO45,DBLP:conf/www/SuchanekKW07} encompasses encyclopedic knowledge up to 2024. Structurally, \icewsa provides highly complete temporal annotations, with temporal structure completeness approaching 100\%, while \wikidata and \yago fall significantly below 50\%. The average temporal density of valid entities also varies dramatically across different TKGs~\cite{WWW25}. Fig.~\ref{introduction3} further illustrates raw data obtained from real-world sources (\wikidata and \yago), which stands in stark contrast to the simplified scenarios shown in Fig.~\ref{introduction2}.
These observations reveal that real-world TKGs are characterized by \textbf{heterogeneous sources with mixed temporal elements} and \textbf{varying structural designs}~\cite{WWW25}. Since the simplified scenarios assumed by current methods cannot reflect real-life challenges, in this work, we propose the TKGA in the wild (TKGA-Wild) task, which underlines two critical challenges in the real world:

\begin{figure*}[t]
	\centering
	\subfigure[Existing simplified TKGA dataset (e.g., \yw).\label{introduction2}]{
		\includegraphics[width=0.45\textwidth]{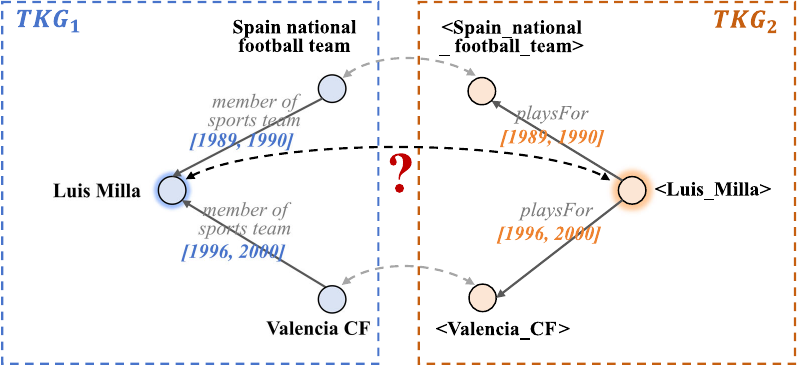}}
	~\quad
	\subfigure[TKGA in the wild.\label{introduction3}]{
		\includegraphics[width=0.451\textwidth]{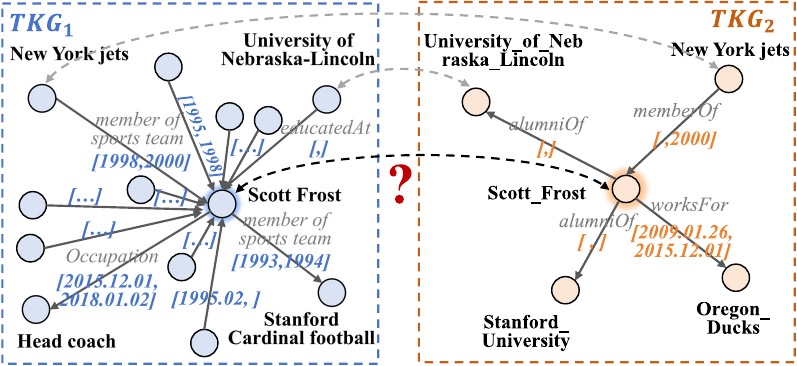}}
    \vspace{-4pt}
	\caption{An illustrative comparison between current simplified TKGA scenarios (a) and real-world TKGA scenarios (b).}
	\label{fig:example}
\vspace{-16pt}
\end{figure*}

\squishlist

\item \emph{\textbf{Multi-Scale Temporal Element Entanglement.}} Real-world TKGs exhibit entangled temporal elements across two critical dimensions:~\emph{\textbf{1) Multi-Granular Temporal Coexistence.}} 
In practice, an entity may simultaneously present temporal elements at varying granularities—ranging from macro-scale patterns (e.g., centuries) to micro-level event records (e.g., months or specific dates). However, existing methods often normalize temporal data to a single granularity (e.g., year-level), discarding information at other important granularities and oversimplifying real-world patterns (e.g., Fig.\ref{introduction2} and Fig.\ref{introduction3}). This \emph{multi-granular temporal coexistence} challenges existing methods~\cite{Zeng2024BenchmarkF,simplehhea,wsdm}, which typically operate at a single temporal granularity and thus struggle to capture such complex patterns; \emph{\textbf{2) Temporal Interval Topological Disparity.}} In real-world scenarios, aligned entities from different TKGs typically exhibit substantial disparities in their overall temporal intervals, manifesting a wide array of topological relationships. As shown in Fig.~\ref{introduction3}, the \emph{Scott Frost} from different TKGs have completely non-overlapping temporal intervals. These relationships include \emph{disjointness} (e.g., one entity spans 2019–2020 while its counterpart covers 1998–2000), \emph{overlap} (e.g., 2019–2024 vs. 2018–2023), \emph{containment} (e.g., 2023 vs. 2020–2024), and \emph{identity} (e.g., 1989–2000 vs. 1989–2000). However, existing methods typically oversimplify this complexity by assuming that aligned entities have identical or highly overlapping temporal intervals, thereby classifying entities with significant temporal discrepancies as misaligned. Such diverse interval topologies significantly complicate temporal correlation reasoning, making it difficult for current methods~\cite{simplehhea,chatea} to accurately model and align entities.

\item \emph{\textbf{Cross-Source Temporal Structural Imbalance.}} TKGA-Wild scenarios face significant structural asymmetries in two key aspects:~\emph{\textbf{1) Multi-Source Temporal Structural Incompleteness.}}  Real-world TKGs frequently exhibit missing or incomplete structural timestamps across different sources. Fig.~\ref{introduction3} highlights such inconsistencies, where neighbor edges for aligned entities simultaneously  lack temporal annotations in both TKGs. However, existing methods predominantly assume most entities are associated with complete temporal relations (time-annotated facts containing both head or tail timestamps, e.g., \emph{member of sports team} in Fig.\ref{introduction2}). This \emph{multi-source temporal incompleteness} leads to asymmetrical structural representations, hindering alignment models~\cite{simplehhea,dualmatch} from learning meaningful correlations;~\emph{\textbf{2) Temporal Event Density Imbalance.}} Additionally, in real-world scenarios, severe asymmetry in temporal fact quantities between aligned entities. For example, in Fig.~\ref{introduction3}, the \emph{Scott Frost} from $TKG_1$ has dozens of timestamped records, while its counterpart has only two. This imbalance in temporal event density poses significant challenges for existing methods~\cite{wsdm,dualmatch}, which typically assume that aligned entities have a roughly similar number of temporal facts (Fig.~\ref{introduction2}). When faced with large discrepancies, these methods often interpret the entities as misaligned. Consequently, the inference of temporal consistency in such real-world alignment settings becomes unreliable.

\squishend

As previously mentioned, existing works neglect these challenges or solve them by naive initialization~\cite{chatea,simplehhea,wsdm}, leading to inadequate utilization of temporal elements and structural information. To address TKGA-Wild challenges, we propose \ourmodel, a new \underline{hy}pergraph-driven \underline{d}ecomposable \underline{r}etrieval-\underline{a}ugmented framework. Technically, we reformulate the TKGA process in a \emph{multi-scale hypergraph retrieval-augmented generation} manner. It first encodes multi-granular temporal, structural, and semantic features to generate similarity matrices and pseudo-aligned pairs, enhancing \emph{multi-granularity temporal understanding}. Then, based on similarity matrices and pseudo-aligned pairs, a \emph{scale-adaptive entity projection} module decomposes and aligns entities across varying temporal and relational scales, constructing a projection hypergraph that captures \emph{complex temporal interval topological disparity} and balances \emph{temporal event density distributions}. This hypergraph, along with similarity cues, is processed by a \emph{multi-scale hypergraph retrieval} module that constructs rich high-order representations, i.e., multi-scale hypergraphs. A \emph{multi-scale interaction-augmented fusion} module then integrates information from multi-scale hypergraphs through \emph{scale-weave synergy mechanisms} (i.e., intra-scale interaction and conflict detection) to infer final entity pairs and effectively resolve issues of \emph{multi-source temporal incompleteness} and \emph{inconsistencies arising from complex and uneven temporal event density distributions}. Through iterative refinement, \ourmodel produces high-quality entity pairs under TKGA-Wild scenarios.

Currently, there is no standard benchmark that captures these challenges of TKGA-Wild and effectively evaluates existing methods (Tab.~\ref{table:introduction}). For example, \dicews is established by splitting the temporal KG \icews into two partially overlapped KG subsets, which weakens the difficulty of TKGA as the KGs to be aligned come from the same source with shared semantics.
\yw originates from two KGs~\cite{DBLP:conf/www/SuchanekKW07,DBLP:conf/semweb/ErxlebenGKMV14}, while it is constructed through manual simplification by selecting quadruples that involve the most frequently occurring and strongly correlated temporal relations (e.g., \emph{playsFor} in Fig.~\ref{introduction2}).
In consequence, the KG pairs to be aligned are in essence the KGs about the football players and their teams, which is limited to a specific domain and cannot reflect the real-life TKGA cases.

\begin{table}[t]
\centering
\caption{Statistics for different TKGA datasets.}
\label{table:introduction}
\vspace{-6pt}
\setlength{\tabcolsep}{3pt}
\resizebox{0.98\columnwidth}{!}{
\begin{tabular}{cbnoi}
\toprule
\rowcolor{white}
& \multicolumn{2}{c}{\textbf{\thead{Multi-Scale Temporal \\Element Entanglement}}} & \multicolumn{2}{c}{\textbf{\thead{Cross-Source Temporal \\Structural Imbalance}}} \\
\cmidrule(lr){2-3}\cmidrule(lr){4-5}
\rowcolor{white} 
\multirow{-4}{*}{\textbf{Dataset}} & \textbf{\thead{Multi-\\Granularity}} & \textbf{\thead{Temporal Interval\\ Topological Disparity}} & \textbf{\thead{Multi-Source \\Incompleteness}} & \textbf{\thead{Temporal Event\\ Density Imbalance}} \\
\midrule
\dicews  & \xmark & \xmark & \xmark & \xmark  \\
\yw & \xmark & \xmark & \xmark  & \xmark \\
\icewswiki & \xmark & \xmark  & \xmark & \cmark \\
\icewsyago & \xmark & \cmark  & \xmark & \cmark \\
\midrule
\Beta (Ours)  & \cmark  & \xmark & $\tmark$ & \xmark \\
\wild (Ours) & \cmark   & \cmark & \cmark & \cmark\\
\bottomrule
\vspace{-13pt}
\end{tabular}
}
\end{table}

To this end, we formally propose to \underline{b}enchmark chall\underline{e}nges for \underline{t}emporal knowledge graph \underline{a}lignment in the \underline{wild} and establish two new datasets, i.e., \Beta and \wild, to better exhibit the difficulty of TKGA-Wild and validate the effectiveness of method. Specifically, we avoid the resource-consuming human labeling by adopting the prior mappings between \yago and \wikidata entities to create the preliminary seeds. 
Then we screen the two KGs and add both temporal and non-temporal facts to establish the aligned KG subsets. 
The new datasets feature \emph{multi-scale temporal element entanglement}, \emph{more realistic cross-source temporal structural imbalance}, and \emph{new challenging alignment scenarios}.

To validate the proposed \ourmodel and TKGA-Wild datasets, we conduct comprehensive experiments on \Beta, \wild, and six representative benchmarks, evaluating against 24 competitive baselines. Results show that \Beta and \wild more faithfully capture real-world TKGA challenges, while leaving room for further advancement. Moreover, \ourmodel establishes a new and effective paradigm for tackling TKGA-Wild and broader alignment tasks, consistently achieving state-of-the-art (SOTA) results and delivering up to a 43.3\% improvement in Hits@1.

\myparagraph{Contribution}This paper extends our prior work~\cite{Zeng2024BenchmarkF} with the following substantial improvements:
\squishlist


\item We expand the scope of TKGA beyond multi-granular temporal information to broader real-world challenges, formally characterizing them as TKGA in the wild (TKGA-Wild), i.e., \emph{multi-scale temporal element entanglement} and \emph{more realistic cross-source temporal structural imbalance}. 

\item We propose a new TKGA-Wild method, \ourmodel, which extends the idea of multi-granular encoding. Our approach for the first time reformulates the TKGA-Wild process in a \emph{multi-scale hypergraph retrieval-augmented generation} manner, and incorporates key modules including \emph{scale-adaptive entity projection}, \emph{multi-scale hypergraph retrieval}, and \emph{multi-scale interaction-augmented fusion}, which collectively address the major challenges of TKGA-Wild.

\item We design \emph{scale-weave synergy} mechanisms for \ourmodel to explicitly address the challenges posed by \emph{multi-source temporal incompleteness} and \emph{temporal event density imbalance}. They effectively integrate intra-scale interactions with cross-scale conflict detection, enabling enhanced dynamic reasoning across complex temporal alignment.

\item We additionally establish a more realistic TKGA-Wild benchmark dataset, \wild, to address the lack of benchmark evaluations for TKGA-Wild.

\item We significantly expand the empirical evaluation by incorporating 14 additional advanced and representative baselines along with 5 influential datasets, thereby enabling a more comprehensive and rigorous analysis. Notably, the performance improves significantly over the initial conference version, with up to 286.0\% improvement on Hits@1 (Tab.~\ref{tab:scenarios}).

\squishend

In summary, the main contributions of this paper are as follow:
\squishlist
\item \underline{\emph{Pioneering Research in TKGA-Wild.}} To the best of our knowledge, this is the first work to formally identify the challenges towards TKGA in the wild and attempt to address them, establishing a foundational approach and benchmark for future explorations in this more realistic area.

\item \underline{\emph{New and Effective TKGA-Wild Framework.}} We propose a new TKGA-Wild framework, \ourmodel, which for the first time efficiently addresses the challenges of TKGA-Wild from a \emph{multi-scale hypergraph retrieval-augmented generation} perspective.

\item \underline{\emph{TKGA-Wild Optimization Techniques.}} To further mitigate \emph{multi-source temporal incompleteness} and \emph{temporal event density imbalance} in TKGA-Wild, we design \emph{scale-weave synergy} mechanisms for \ourmodel. Specifically, the mechanisms incorporate both \emph{intra-scale interactions} and cross-scale \emph{conflict detection}, enabling explicit modeling of dynamic relationships and the resolution of inconsistencies across temporal scales.

\item \underline{\emph{New TKGA-Wild Benchmarks.}} To address the lack of benchmark evaluations for TKGA-Wild, we formally propose to benchmark challenges for TKGA-Wild and validate the effectiveness of the method by establishing two new datasets, i.e., \Beta and \wild to inspire follow-up research. 

\item \underline{\emph{Comprehensive Validation Through Extensive Experiments.}} We conduct extensive experiments on \Beta, \wild, and current six datasets using \ourmodel and 24 SOTA solutions. The Results not only validate the difficulty of the TKGA-Wild datasets but also demonstrate the superior effectiveness of \ourmodel, achieving up to a 43.3\% improvement in Hits@1.

\squishend







\section{Related Work}
\label{sec:background}

While KGA methods and its benchmarks have been well studied~\cite{chatea}, only a few works explore the TKGA methods and its benchmarks. We review prior work on KGA, TKGA, retrieval-augmented generation, and TKGA datasets.

\myparagraph{Knowledge Graph Alignment}
Traditional knowledge graph alignment has been intensively studied over the past few years~\cite{ICDEea01,ICDEea02,ICDEea03,zhaosurvey22,DBLP:books/sp/ZhaoZT23,DBLP:journals/pvldb/SunZHWCAL20,DBLP:conf/cikm/YangWZQWHH21,DBLP:journals/vldb/ZhangTLJQ22,AAAI25EA,TKDE25EA,WSDM25EA}.
Early research primarily focused on translation-based methods~\cite{MTransE,BootEA,transE,HybEA24,sun20}, which aim to learn cross-KG representations and infer entity equivalence within a shared embedding space~\cite{transE,MTransE,DBLP:conf/semweb/SunHL17,DBLP:conf/sigir/GeLCZG21}. However, they often struggle to capture complex structural and semantic contexts. With the advancement of deep graph neural networks (GNNs), recent studies have shifted toward GNN-based~\cite{GCN-Align,RDGCN,Dual-AMN,mraea,rrea} and other emerging methods~\cite{Fualign,BERT-INT,llm4ea,twoea,bert,PARIS}, which focus on improving accurate alignment inference~\cite{DBLP:journals/tkde/ZengZTTC23,ICDEea05}, handling large-scale data~\cite{DBLP:conf/cikm/LiuHZZZ22,DBLP:conf/kdd/GaoLW0W022,VLDBea01}, and operating under low-resource supervision settings~\cite{DBLP:conf/cikm/Zeng0TF21,DBLP:conf/cikm/ZengD00HYLCWLHF22,WSDM25EA,AAAI25EA}. Nevertheless, methods designed for general KGA tasks are not well-suited for TKGA, as they fail to effectively incorporate and reason over temporal information.

\myparagraph{Temporal Knowledge Graph Alignment} As an emerging research area, there are still only a limited number of studies dedicated to temporal knowledge graph alignment~\cite{Zeng2024BenchmarkF,wsdm,nn25,CTEA,TKGARP}. Most existing methods extend traditional static knowledge graph alignment approaches by incorporating temporal encoding to improve alignment performance. Notably, \TEAGNN and \TREA are among the first to explore the TKGA task~\cite{TEA-GNN,TREA}. Unlike temporal GNN models that discretize temporal graphs into multiple snapshots, these models treat timestamps as attributes of the links between entities. Tem-EA employs a long short-term memory (LSTM) network to encode temporal sequences, which are then combined with structural embeddings learned via GCN to produce alignment results~\cite{Tem-EA}. TGA-EA enhances graph attention mechanisms with temporal modeling to learn temporal-relational entity embeddings for alignment~\cite{TGA-EA}. \TEA proposes a time-aware entity alignment framework that captures entity evolution by leveraging temporal context and aggregating diverse contextual information for alignment decisions~\cite{TEA}. For unsupervised TKGA, recent methods such as \dualmatch and \STEA leverage temporal information without relying on labeled data, using simple architectures and new matching strategies to align entities effectively~\cite{dualmatch,STEA}. Beyond GNN-based methods, \lightea adopts a two-aspect, three-view label propagation strategy to infer entity labels~\cite{lightea}. \SimpleHHEA addresses heterogeneity in temporal graphs by leveraging auxiliary information~\cite{simplehhea}. \ChatEA explores a new direction by employing fine-tuned large language models for entity alignment, which can also be applied to TKGA tasks~\cite{chatea}.

Unlike existing approaches, which fail to effectively address the challenges of TKGA-Wild, this work is the first to formulate the TKGA-Wild task as a \emph{multi-scale hypergraph retrieval-augmented generation} paradigm. We focus on modeling key issues that have been largely overlooked, including \emph{multi-granular temporal information}, \emph{temporal interval topological disparity}, \emph{multi-source temporal incompleteness-induced complex alignment patterns}, and \emph{temporal event density imbalance distributions}.

\myparagraph{Retrieval-Augmented Generation}
Retrieval-augmented generation (RAG) has garnered significant attention across various domains in recent years due to its effectiveness in capturing information and enhancing the performance of models on knowledge-intensive tasks~\cite{TPAMI3,selfrag,ragsurvey2,GuuLTPC20,IzacardCHRBJG22,ragaaai25,ragwww25,agentrag,zhang2025rakg,TPAMI5,ragcoling25,selfcot,naive,ICLRRAG25}. For example, Self-RAG~\cite{selfrag} improves the generation quality of large language models (LLMs) by retrieving external information on demand and incorporating self-reflection. RAKG~\cite{zhang2025rakg} organizes training data into a retrieval-augmented manner, enabling the retrieval of relevant textual and graphical content to improve knowledge graph construction. RAGTrans~\cite{ragKDD24} stores training data in a multimodal memory for knowledge retrieval and enhances user content generation by aggregating neighborhood information through hypergraph transformers on a multimodal hypergraph. Inspired by this paradigm, we propose a new TKGA-Wild framework, which for the first time applies the RAG concept to TKGA. Our approach introduces a multi-scale hypergraph paradigm that effectively captures high-dimensional temporal alignment patterns both before and after retrieval, addressing the challenges posed by the TKGA-Wild task.


\myparagraph{Temporal Knowledge Graph Alignment Datasets}
Currently, the most widely used datasets, \dicews and \yw, are extracted from \icewsa, \yago and \wikidata. 
\icewsa is originally extracted from \icews that contains political events with specific time points, with time unit of 1 day, and  
\icewsa contains events during 2005 to 2015.
Subsequently, the quadruples in \icewsa are randomly divided into two subsets of approximately equal size, resulting in the formation of two datasets: \da (shortened as \das) and \db (shortened as \dbs).
The difference is the number of seeds, i.e., 200 for \das and 1000 for \dbs. 

The \yw datasets are built upon \wikidata and \yago~\cite{TEA-GNN}. Specifically, the top 50,000 most frequent entities are first selected from a subset of Wikidata extracted by the work~\cite{DBLP:conf/iclr/LacroixOU20}, and then these entities are linked to their counterparts in YAGO.
Temporal facts are subsequently added to form two TKGs for alignment.
\ywa (shortened as \ywas) contains 1000 seed pairs, while \ywb (shortened as \ywbs) contains 5000. 
\ywc is also constructed by the work~\cite{TEA-GNN}.
A preliminary experiment on \ywc unveils that the model designed specifically for temporal entity alignment performs worse on non-temporal facts~\cite{TEA-GNN}. 
Notably, in these datasets, only the year information is retained to ensure shared time set. Similarly, \icewswiki and \icewsyago, as newly proposed heterogeneous KGA datasets with temporal information, also simplify time representation to a single granularity~\cite{simplehhea}.
The statistics of the mainstream datasets are provided in Tab.~\ref{tab:stats}. 

Hence, current TKGA datasets, either obtained from one single KG, or unified temporal element standards and simplified temporal structures, 
may not serve as a fair benchmark in terms of evaluating TKGA performance in the wild.


\section{Methodology}\label{sec:method}

To address the aforementioned issues, in this section, we present \ourmodel, a hypergraph-driven decomposable retrieval-augmented framework, enabling multi-scale knowledge fusion for TKGA-Wild. We first formally define the problem in Section~\ref{sec3:definition}. Then, we introduce the overall framework in Section~\ref{sec3:overview}, followed by details of its key modules in Section~\ref{sec3:granular}-\ref{sec3:fusion}.

\begin{figure*}[ht]
	\includegraphics[width=0.98\textwidth]{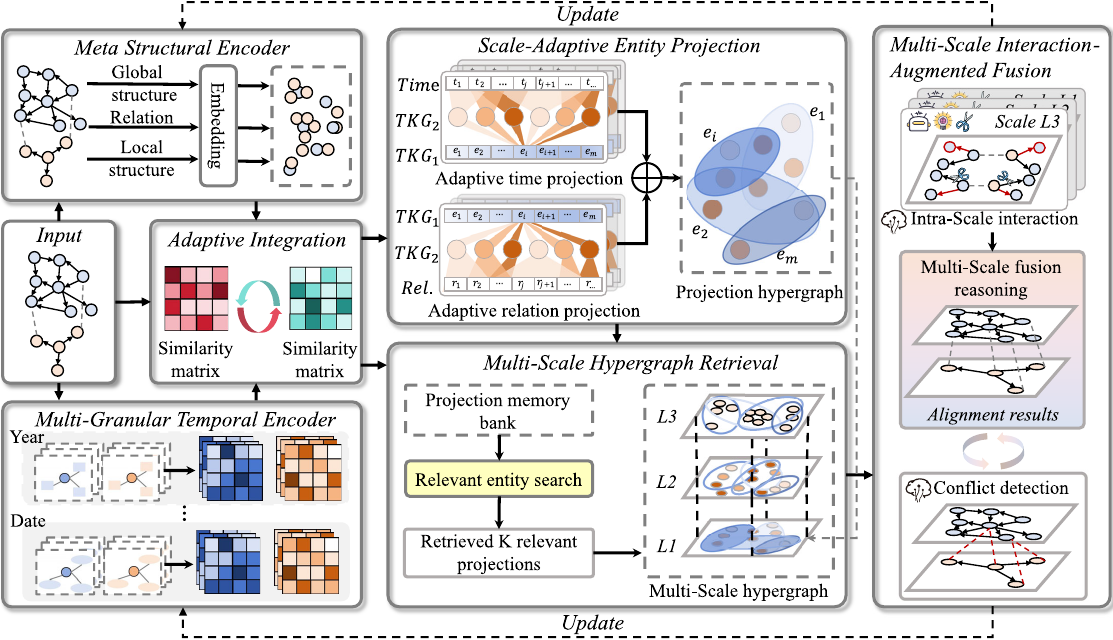}
	\caption{The overall framework of \ourmodel.}
	\label{framework}
\end{figure*}

\begin{table}[htbp]
  \centering
  \caption{Main notations used in this paper.}
  \renewcommand{\arraystretch}{1.1} 
  \begin{tabular}{ll}
    \toprule
    Notation & Meaning \\
    \midrule
    $PATH_n$ & Random walk path used in structural encoding \\
    $\mathcal{N}_{e_i}^{-}$ & 1-hop neighbors of $e_i$ excluding $e_{i-1}$ \\
    $\mathbf{t}_n^{(g)}$ & Binary time vector of entities (granularity $g$) \\
    $\textbf{t2v}(t)$ & Time2Vec embedding for timestamp $t$ \\
    $\boldsymbol{h}_n^{(g)}$ & Temporal embedding of entities (granularity $g$) \\
    $\boldsymbol{h}^{dw}_n, \boldsymbol{h}^{\text{name}}_n$ & Structural/Name-based embedding of entities \\
    $\boldsymbol{h}^{ada}_n$ & Adaptively integrated embedding of entities \\
    $\boldsymbol{P}^{ada}$ & Adaptive similarity matrix between TKGs \\
    $\mathbb{P}_{time}(\cdot), \mathbb{P}_{rel}(\cdot)$ & Time/Relation projection operator\\
    $\tilde{\mathbb{P}}_{time}^{e^{s,t}}, \tilde{\mathbb{P}}_{rel}^{e^{s,t}}$ & Projected target entity timestamps/relations for $e^s$ \\
    $\mathcal{G}_p = (\mathcal{V}_p, \mathcal{E}_p)$ & Projection hypergraph: nodes and hyperedges \\
    $\mathcal{M}_p$ & Projection memory bank (vector database) \\
    $\phi_\ell$ & Aligned entity pairs at scale $\ell$ \\
    $\phi_C$ & Conflict-resolved entity pairs across scales \\
    $\phi_{f}$ & Final fused alignment result\\
    $I_n$ & Number of iterative update rounds \\
    \bottomrule
  \end{tabular}
  \label{tab:notation}
\end{table}

\subsection{Task Definition}\label{sec3:definition}

\begin{definition}
A \textbf{temporal knowledge graph} (TKG) can be denoted as $G = (E,R,T,Q)$, where $E$ is the set of entities, $R$ is the set of relations, $T$ is the set of time intervals, and $Q=\{(e_h,r,e_t, \tau)~|~e_h,e_t \in E, r \in R, \tau \in T \}$ is the set of fact quadruples, each of which represents that the subject entity $e_h$ has the relation $r$ with the object entity $e_t$  during the time interval $\tau$. $\tau$ is represented as $[t_b, t_e]$,  with the beginning timestamp $t_b$ and the ending timestamp $t_e$. If $\tau$ is not none, the fact is referred to as a \emph{valid quadruple} or a \emph{temporal fact}.

\end{definition}

\begin{definition}
\label{sec:problem_statement}
\textbf{Temporal knowledge graph alignment in the wild} (TKGA-Wild) aims to find equivalent entities $\phi$ from a source TKG\,$G^s = (E^s,R^s,T^s,Q^s)$ to a target TKG $G^t = (E^t,R^t,T^t, Q^t)$. $\phi = \{(e^s, e^t) \in E^s \times E^t~|~e^s \Leftrightarrow e^t\}$ , where $e^s \in E^s$, $e^t \in E^t$, and $\Leftrightarrow$ is an equivalence relation between two entities. The main notations used are presented in Tab.~\ref{tab:notation}.
\end{definition}

\subsection{Framework Overview}\label{sec3:overview}

The overall framework is presented in Fig.~\ref{framework}. \ourmodel treats the problem as a \emph{multi-scale hypergraph retrieval-augmented generation} process, addressing the challenge of TKGA-Wild through a multi-scale hypergraph decomposition approach. This method effectively tackles issues such as \emph{multi-granular temporal coexistence}, \emph{temporal interval topological disparity}, \emph{multi-source temporal incompleteness}, and \emph{temporal event density imbalance distributions}.

\ourmodel first utilizes two key encoding modules for feature extraction: the \emph{meta structural encoder} module generates entity representations through the local and global structure of the TKGs, while the \emph{multi-granular temporal encoder} module focuses on modeling multi-granularity temporal information. The features learned by both modules are then passed to the \emph{adaptive integration} module, which generates a adaptive similarity matrix and produce pseudo-aligned entity pairs.

Next, the pseudo-entity pairs and the adaptively fused similarity matrix are passed into the \emph{scale-adaptive entity projection} module. In this module, entities are adaptively aligned based on temporal and relational features at different scales, generating more informative scale-adaptive aligned projections, which are then fused to form a projection hypergraph.

Subsequently, the adaptively fused similarity matrix, pseudo-entity pairs, and projection hypergraph are fed into the \emph{multi-scale hypergraph retrieval} module. This module stores the information in the projection memory bank and retrieves relevant entities based on this bank, generating a multi-scale hypergraph with higher-order information representation. These multi-scale hypergraphs are then passed into the \emph{multi-scale interaction-augmented fusion} module, which performs inference by \emph{scale-weave synergy mechanisms} (i.e., \emph{intra-scale interaction} and \emph{conflict detection}), thus fusing temporal information across different scales to derive the alignment result, which is then fed back to update the two key encoding modules.

Through multiple rounds of iterative fusion and multi-scale hypergraph retrieval, \ourmodel effectively overcomes multiple challenges in TKGA-Wild, ultimately generating high-quality alignment results.


\subsection{Multi-Granular Information Encoders and Integration}\label{sec3:granular}
To effectively support the downstream multi-scale retrieval and alignment processes, it is essential to provide rich, fine-grained, and semantically coherent entity representations. Therefore, we begin by introducing a set of \emph{multi-granular information encoders} that model the structural and temporal contexts of entities from complementary perspectives. These encoders lay the foundation for resolving the core challenges of TKGA-Wild (i.e., \emph{multi-granular temporal coexistence}) by capturing both static topological patterns and dynamic temporal behaviors. The resulting multi-view embeddings are then integrated through an \emph{adaptive integration} module, generating an initial alignment signal that bootstraps the entire alignment framework.

\myparagraph{Meta Structural Encoder} Firstly, following previous work~\cite{Zeng2024BenchmarkF,simplehhea}, we also introduce a \emph{meta structural encoder} to effectively capture both local and global structural patterns in TKGs. This encoder combines biased random walks with a skip-gram-based embedding framework and doesn't require supervision~\cite{simplehhea}. Specifically, the random walk process maintains relation information by generating edge-labeled paths of the form ${PATH}_n = \left(e_1, r_1, e_2, \ldots, r_{l-1}, e_l\right)$, where transition probability is controlled by a hyperparameter $\beta$ to balance between breadth-first and depth-first search behaviors. The probability of an entity being selected is defined as:

\begin{equation}
\resizebox{0.9\hsize}{!}{$\begin{aligned}
\operatorname{P}_r\left(e_{i+1} \mid e_i\right)=
\begin{cases}
\beta, & d\left(e_{i-1}, e_{i+1}\right)=2 \\
1-\beta, & d\left(e_{i-1}, e_{i+1}\right)=1
\end{cases}, \quad e_{i+1} \in \mathcal{N}_{e_i}^{-},
\end{aligned}$}
\end{equation}
where $\mathcal{N}_{e_i}^{-}$ denotes the 1-hop neighbors of $e_i$ excluding $e_{i-1}$. Each selected transition also retrieves the corresponding relation $r_i$ from the quadruple $(e_i, r_i, e_{i+1}, \tau_i) \in Q$. The resulting paths are treated as sentences, with entities and relations as tokens. We then apply the Skip-gram model followed by a linear transformation $W_\mathcal{D}$ to learn structural embeddings $\{\boldsymbol{h}^{dw}_n\}_{n=1}^N$ for entities and relations. This process effectively encodes the key structural information in the TKGs.

\myparagraph{Multi-Granular Temporal Encoder}
The \emph{Multi-Granular Temporal Encoder} is designed to capture temporal patterns of entities at different granularities (i.e., year, month, and date) to enhance TKGA by modeling both coarse-grained and fine-grained temporal correlations across a wider range. Inspired by prior work~\cite{kazemi2019time2vec,Zeng2024BenchmarkF,simplehhea}, we adopt the \textit{Time2Vec} to encode temporal features across multiple scales.

For each entity, we annotate its time occurrences based on the timestamps of associated facts in the TKG. In \wild, the time span $T$ covers intervals from 1995 to 2021. We decompose time into three granularities: year, month, and date. Year granularity representations aim to capture broad temporal trends and similarities among entities, while month and date granularity representations focus on fine-grained temporal alignment.

Each granularity yields a binary temporal vector for an entity $e_n$, denoted as $\mathbf{t}_n^{(g)} = \{\mathbf{t}_n^{(g), i}\}_{i=1}^{|T_g|}$, where $g \in \{\text{year}, \text{month}, \text{date}\}$ and $\mathbf{t}_n^{(g), i} = 1$, if $e_n$ is involved in any fact at the $i^{th}$ time point under granularity $g$, and $0$ otherwise.

For each active timestamp $t$ in $\mathbf{t}_n^{(g)}$, we compute its \textit{Time2Vec} encoding as follows:
\begin{equation}
\textbf{t2v}(t)[i] = 
\begin{cases}
\omega_i t + \varphi_i, & i = 0 \\
\cos(\omega_i t + \varphi_i), & 1 \leq i \leq k
\end{cases},
\end{equation}
where $\omega_i$ and $\varphi_i$ are learnable parameters. This formulation allows the model to capture both linear and periodic temporal patterns.

The final time embedding for entity $e_n$ under granularity $g$ is obtained by aggregating the \textit{Time2Vec} representations of all its active timestamps and applying a linear transformation:
\begin{equation}
\boldsymbol{h}_n^{(g)} = W_{\mathcal{T}_g} \cdot \left(\frac{1}{|\mathbb{T}_n^{(g)}|} \sum_{t \in \mathbb{T}_n^{(g)}} \textbf{t2v}(t) \right),
\end{equation}
where $\mathbb{T}_n^{(g)}$ denotes the set of time points where $\mathbf{t}_n^{(g), i} = 1$ and $W_{\mathcal{T}_g}$ is a trainable projection matrix for granularity $g$.

The encoder thus produces a set of multi-granular temporal embeddings $\{\boldsymbol{h}_n^{(\text{year})}, \boldsymbol{h}_n^{(\text{month})}, \boldsymbol{h}_n^{(\text{date})}\}$ for each entity, which are further integrated into downstream TKGA tasks.

\myparagraph{Adaptive Integration} Finally, we generate high-quality embeddings for entities through the \emph{adaptive integration} of different types of embeddings as follow: 
\begin{equation}
\resizebox{0.9\hsize}{!}{$\begin{aligned}
\{\boldsymbol{h}^{ada}_n\}_{n=1}^N = \{[\boldsymbol{h}^{name}_n \otimes \boldsymbol{h}^{(year)}_n \otimes \boldsymbol{h}^{(month)}_n \otimes \boldsymbol{h}^{(date)}_n \otimes \boldsymbol{h}^{dw}_n]\}_{n=1}^N,
\end{aligned}$}
\end{equation}
Notably, the name-based embeddings ${\{\boldsymbol{h}^{\text{name}}_n}\}_{n=1}^N$ 
are obtained in an unsupervised manner using a previously proposed BERT-based dual transformation method~\cite{simplehhea}, which efficiently encodes entity names into the same dimensional space as the other embedding types. This design reflects our intention to further incorporate semantic information from multimodal sources to enhance the quality of the embeddings.

We adopt the margin ranking loss as the training objective for TKGA inference. During testing, we employ the cross-domain similarity local scaling~\cite{conneau2017word} as the distance metric to evaluate similarities between source and target TKGs' entity embeddings, resulting in an adaptive similarity matrix between entities across TKGs. Based on this adaptive similarity matrix $\boldsymbol{P}^{ada}$, for each source entity, we retrieve the target entity with the highest similarity score and regard it as the pseudo-aligned pair.


\subsection{Scale-Adaptive Entity Projection}\label{sec3:scale}
Although the fusion similarity matrix obtained through \emph{Adaptive Integration} can initially identify the similarity relationships between temporal entities at different granularities, this simple multi-granularity time information embedding fails to comprehensively represent the true correlations due to the challenges posed by the \emph{complex temporal interval
topological disparity} and the \emph{temporal event density
distributions} in the TKGA-Wild scenarios (this phenomenon can be validated through the experimental results of previous work \mgtea~\cite{Zeng2024BenchmarkF}). To tackle the above dilemmas, we further propose a \emph{scale-adaptive entity projection} module: by deeply decoupling the adaptive similarity matrix $\boldsymbol{P^{ada}}$, each temporal entity is further decomposed along the multiple temporal and relational dimensions into multiple projections with scale-awareness capability, and a projection hypergraph is constructed, thereby enabling the adaptive representation of temporal entities at different scales.

\myparagraph{Adaptive Time \& Relation Projection} The core of this component lies in establishing a fine-grained temporal alignment mapping mechanism. For each source entity's top-$k$ similar target entities in the similarity matrix, we design a dual-channel projection operator: in the time dimension, we apply a timestamp masking projection $\mathbb{P}_{time}(\cdot) $ to remove timestamps in the target entity that do not appear in the source entity as well as the facts attached to their timestamps; in the relation dimension, we implement a relation type masking projection $\mathbb{P}_{rel}(\cdot)$ to eliminate relation types not present in the source entity as well as the facts attached to their relation types. Specifically, given a source entity $e^s \in E^s$ and a target entity $e^t \in E^t$, their projection process can be formalized as:
\begin{equation}
\begin{aligned}
\tilde{\mathbb{P}}_{time}^{e^{s,t}} &= \mathbb{P}_{time}(e^t|\mathcal{T}(e^s)) = \text{Mask}(\mathcal{T}(e^t), \mathcal{T}(e^s)), \\
\tilde{\mathbb{P}}_{rel}^{e^{s,t}} &= \mathbb{P}_{rel}(e^t|\mathcal{R}(e^s)) = \text{Mask}(\mathcal{R}(e^t), \mathcal{R}(e^s)),
\end{aligned}
\end{equation}
where \( \mathcal{T}(\cdot) \) and \( \mathcal{R}(\cdot) \) represent the sets of timestamps and relation types associated with an entity, respectively. Through this dual-channel selective projection, each target entity \( e^t \) corresponding to a similar source entity \( e^s \) generates two projections  $\{\tilde{\mathbb{P}}_{time}^{e^{s,t}}, \tilde{\mathbb{P}}_{rel}^{e^{s,t}}\}$, each maintaining temporal and relational consistency, ultimately forming a projection set of size $|E^s| \times k \times 2$, where $|E^s|$ denote the number of source entities. This decoupled projection effectively addresses the semantic and temporal drift problem in cross-scale TKGA-Wild while preserving the distinguishable features of entities across different dimensions.




\myparagraph{Projection Hypergraph Construction} Based on the above set of projections, we construct a dynamic projection hypergraph $\mathcal{G}_p = (\mathcal{V}_p, \mathcal{E}_p)$. The hypernode set $\mathcal{V}_p$ is constructed to encompass both the original target entities and their corresponding multi-dimensional projections:
\begin{equation}
\begin{aligned}
\mathcal{V}_p = E^t \cup \bigcup_{e^s \in E^s} \bigcup_{e^t \in \text{Top-$k$}(e^s)} \{\tilde{\mathbb{P}}_{\text{time}}^{e^{s,t}}, \tilde{\mathbb{P}}_{\text{rel}}^{e^{s,t}}\},
\end{aligned}
\end{equation}
where $\text{Top-$k$}(e^s) = \arg\max_{k} \{\mathbf{P}^{\text{ada}}[i,j] : e^s_i = e^s, e^t_j \in E^t\}$ denotes the set of $k$ most similar target entities for source entity $e^s$. The hyperedge set $\mathcal{E}_p$ is dynamically generated to establish connections between source entities and their most relevant target entities along with their corresponding projections. Each hyperedge $\mathcal{H}^p_{e^s} \in \mathcal{E}_p$ for a source entity $e^s$ is defined as:
\begin{equation}
\begin{aligned}
\mathcal{H}^p_{e^s} = \{e^s\} \cup \text{Top-k}(e^s) \cup \bigcup_{e^t \in \text{Top-k}(e^s)} \{\tilde{\mathbb{P}}_{\text{time}}^{e^{s,t}}, \tilde{\mathbb{P}}_{\text{rel}}^{e^{s,t}}\}.
\end{aligned}
\end{equation}

The complete hyperedge set is thus formulated as:
\begin{equation}
\begin{aligned}
\mathcal{E}_p = \{\mathcal{H}^p_{e^s} : e^s \in E^s\}.
\end{aligned}
\end{equation}

This design endows the hypergraph with two key properties: 1) by leveraging multi-dimensional projections, it captures multi-scale temporal interaction patterns, enabling the model to differentiate and align asynchronous or uneven temporal intervals—thus explicitly modeling complex temporal interval topological disparity; and 2) by incorporating relation-masked projections, it can adaptively capture genuine semantic associations across different relation types. Notably, the hypergraph dynamically updates in accordance with changes in the adaptive fusion matrix, enabling progressive modeling of complex temporal dependencies among entities. This mechanism effectively addresses the representational rigidity inherent in traditional static graph models when dealing with temporal event density complex and heterogeneous distributions, thereby providing TKGA-Wild with a more flexible representational space.


\begin{figure}[t]
	\includegraphics[width=0.98\linewidth]{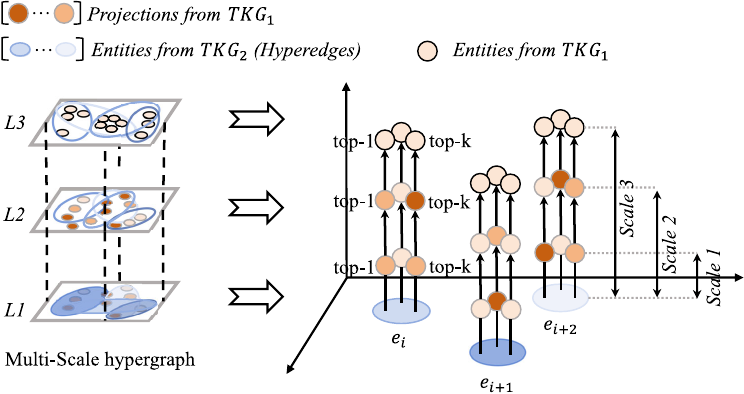}
	\caption{An example of the multi-scale hypergraph.}
	\label{scale}
\end{figure}

\subsection{Multi-Scale Hypergraph Retrieval}\label{sec3:hyper}

We expect to retrieve projections from the target TKG that exhibit genuine relevance and temporal scale alignment, thereby eliminating the interference of irrelevant and misleading temporal information, which in turn contributes to the realization of TKGA-Wild. Inspired by the information aggregation mechanisms in retrieval-augmented generation techniques~\cite{ziyang24}, we reformulates the TKGA-Wild task as a retrieval-augmented generation problem and proposes a \emph{multi-scale hypergraph retrieval} module. This module performs relevant entity retrieval by constructing a projection memory bank, dynamically selecting the top-$k$ most relevant entities, and ultimately integrating them to generate a multi-scale hypergraph structure for \emph{more efficient representation of complex temporal interval topological disparity and temporal event density distributions}.



\myparagraph{Projection Memory Bank and Search}
First, we construct a projection memory bank $\mathcal{M}_p$\footnote{stored by the vector database, \url{https://github.com/facebookresearch/faiss}} that contains the target entities along with their projection information. The projection information encompasses various aspects of target entity features, such as time and relation information at different scales, aiming to reflect the multi-dimensional characteristics of the target entity. This enables a more fine-grained relevance assessment. In order to efficiently utilize this resource, we design a relevant entity search module, which retrieves the top-$k$ most relevant projection information from the memory bank for each source entity $e^s \in E^s$:
\begin{equation}
\{\tilde{\mathbb{P}}_{1}, \tilde{\mathbb{P}}_{2}, \ldots, \tilde{\mathbb{P}}_{k}\}_{i=1}^{|E^s|} = \arg\max_{k} \text{Retrieval}(E^s, \mathcal{M}_p),
\end{equation}
where $\text{Retrieval}(\cdot,\cdot)$ represents the similarity-based (e.g., cosine similarity) retrieval function.

\myparagraph{Multi-Scale Hypergraph} Subsequently, to effectively integrate the information across different scales, we construct a multi-scale hypergraph by combining the projection hypergraph $\mathcal{G}_p$, the retrieved top-$k$ projections, and abstraction projection fusion. As shown in Fig.~\ref{scale}, this structure captures higher-dimensional and more fine-grained temporal information representations. Specifically, the constructed multi-scale hypergraph $\mathcal{G}_m = (\mathcal{V}_m, \mathcal{E}_m)$ consists of three hierarchical layers:


\squishlist
\item \emph{Layer 1 (L1, $\ell = 1$)} which captures the finest details, consists of a projection hypergraph $\mathcal{G}_p = \mathcal{G}_1 = (\mathcal{V}_1, \mathcal{E}_1)$ generated by the \emph{scale-adaptive entity projection} module. This hypergraph primarily derives from the \emph{meta structural encoder} and \emph{multi-granular temporal encoder}, as well as the similarity matrix $\boldsymbol{P}^{ada}$ produced by the \emph{adaptive integration}, incorporating \emph{scale-adaptive entity projection}. It is capable of initially capturing effective temporal information under \emph{multi-granularity temporal coexistence}, \emph{temporal interval topological disparity} and \emph{temporal event density imbalance}.

\item \emph{Layer 2 (L2, $\ell = 2$)} constructs a hypergraph $\mathcal{G}_2 = (\mathcal{V}_2, \mathcal{E}_2)$ based on retrieved projections, where $\mathcal{V}_2 = \arg\max_{k} \text{Retrieval}(E^s, \mathcal{M}_p)$, $\mathcal{E}_2 = \{\mathcal{H}^2_{e^s_i} = \arg\max_{k} \text{Retrieval}(e^s_i, \mathcal{M}_p) \mid e^s_i \in E^s\}$. Each hypernode in $\mathcal{V}_2$ represents a retrieved relevant projection $\tilde{\mathbb{P}}$ retrieved from the projection memory bank $\mathcal{M}_p$, and each hyperedge in $\mathcal{E}_2$ represents a query (i.e., the source entity), connecting all retrieved relevant projections corresponding to that source entity. By using a RAG perspective, multi-scale information is reorganized to further capture the more relevant temporal information.

\item \emph{Layer 3 (L3, $\ell = 3$)} performs abstract fusion and simplification, forming a hypergraph $\mathcal{G}_3 = (\mathcal{V}_3, \mathcal{E}_3)$ that combines and compresses information from \emph{L2} into a more abstract form of hypergraph:
\begin{align}
\mathcal{V}_3 &= \{\phi(\tilde{\mathbb{P}}) \mid \tilde{\mathbb{P}} \in \mathcal{V}_2\}, \\
\mathcal{E}_3 &= \{\mathcal{H}^3_{e^s_i} = \{\phi(\tilde{\mathbb{P}}) \mid \tilde{\mathbb{P}} \in \mathcal{H}^2_{e^s_i}\} \mid e^s_i \in E^s\},
\end{align}
where $\phi(\tilde{\mathbb{P}})$ maps the retrieved projection to its corresponding target entity. This further compresses the temporal information and focuses on temporal alignment from a high-level perspective.

\squishend

This hierarchical design enables the multi-scale hypergraph to simultaneously preserve low-level temporal details (derived from multi-granular and extensive, scale-adaptive projections, i.e., \emph{L1}) and abstract high-level alignments (derived from relevant projection retrieval and abstraction fusion, i.e., \emph{L2 and L3}), resulting in a more expressive and adaptable representation of source-target entity relationships.
Formally, the multi-scale hypergraph $\mathcal{G}_m$ is structured as:
\begin{equation}
\mathcal{G}_m = \left(\bigcup_{\ell=1}^{3} \mathcal{V}_{\ell}, \bigcup_{\ell=1}^{3} \mathcal{E}_{\ell}\right).
\end{equation}

This multi-scale structure allows for a dynamic adjustment of temporal dependencies between entities and ensures flexibility in the alignment process.

\subsection{Multi-Scale Interaction-Augmented Fusion}\label{sec3:fusion}
To facilitate more comprehensive temporal alignment paradigm interactions and information utilization across different scales, a direct approach is to mix entity feature information from various scales. However, we argue that intra-scale and inter-scale interactions represent different dimensions of temporal feature pattern interactions. Specifically, intra-scale interactions primarily describe detailed interactions between eat similar time scales, while inter-scale interactions focus on the macroscopic changes and conflicts in entity features~\cite{timemixer,adaMSHyper}. Therefore, instead of directly mixing multi-scale temporal entity information as a whole, we propose a \emph{multi-scale interaction-augmented fusion} module. This module addresses the \emph{multi-source temporal incompleteness} and \emph{temporal event density imbalance} challenges in TKGA-Wild through two main components: \emph{scale-weave synergy mechanisms} and \emph{multi-scale fusion reasoning}.
The \emph{scale-weave synergy mechanisms} integrate both \emph{intra-scale interactions} and cross-scale \emph{conflict detection}, enabling scale-adaptive supplementation and pruning of temporal incomplete or imbalanced data. This facilitates effective regulation of temporal event density and resolution of inconsistencies across temporal scales. On top of this, the \emph{multi-scale fusion reasoning} component adaptively aggregates the interaction-enhanced representations from multiple scales, promoting more robust and coherent temporal feature understanding.

\myparagraph{Intra-Scale Interaction}
Due to the issue of incomplete temporal information in TKGA-Wild, traditional alignment models may lead to information bottlenecks~\cite{timemixer}. Recent studies~\cite{timemixer,adaMSHyper} have shown that group-wise interactions can provide richer insights for incomplete temporal information. To align entities with similar temporal information within each scale, we introduce time fact supplementation and time fact trimming operations within the \emph{intra-scale interaction mechanism}. Specifically, given a particular scale $\ell$, for each pair of similar entities in the hypergraph $\mathcal{G}_\ell$, the time fact supplementation operation allows the LLM to analyze the entity pair's situation and supplement the missing core temporal information (including time, relations, or facts). The time fact trimming operation allows the LLM to trim away irrelevant time information, i.e., $\mathcal{I}_\ell: \mathcal{G}_\ell \rightarrow \mathcal{G}_\ell^{*}$, where $\mathcal{G}_\ell^{*}$ represents the enhanced hypergraph after supplementation and trimming operations. Applying this process across all scales produces the updated multi-scale hypergraph $\mathcal{G}^{*}_m$.

\myparagraph{Multi-Scale Fusion Reasoning}
To reduce the search space in fusion reasoning and mitigate hallucination phenomena~\cite{Zhang24, kgllmsurvey24, llmsurveyyang24}, we propose a \emph{multi-scale fusion reasoning}. This component transforms the temporal alignment problem across different scales into a selection task. Specifically, the component handles two types of multi-scale temporal information inputs: 1) the updated multi-scale hypergraph $\mathcal{G}^{*}_m$ from the \emph{intra-scale interaction}; and 2) the conflicting sets $\{C(e^s_i)\}_{i=1}^{|E^s|}$ from \emph{conflict detection}. For the first input, the component combines similar temporal entity pairs from the same source entity and same scale. This converts the problem into a selection task for each source entity, where the LLM selects relevant temporal entity pairs from the corresponding entities, generating the aligned entity pairs for each scale. The result is denoted as $\phi_\ell = \{(e^s_i, e^t_j) \mid e^s_i \in E^s, e^t_j \in E^t\}$, where $\phi_\ell$ represents the aligned entity pairs after selection and alignment at scale $\ell$. 

For the second input, the component transforms the problem into a cross-scale selection task where the LLM selects relevant temporal entity pairs for each source entity from conflicting temporal entity pairs across all scales. This process generates conflict-resolved entity pairs $\phi_C = \{(e^s_i, e^t_j) | e^s_i \in E^s, e^t_j \in E^t\}$, which represent the optimal temporal alignment choices after resolving inter-scale conflicts. This fusion approach and the scale-weave synergy mechanisms (\emph{intra-scale interaction} and \emph{conflict detection}) ensures coherent temporal reasoning while maintaining computational efficiency through search space reduction.

\myparagraph{Conflict Detection}
Considering that relying solely on \emph{intra-scale interaction} may cause conflicts across scales, which could impact the efficiency of \emph{multi-scale fusion reasoning}, a direct approach to addressing this is to first model all entity interactions across all scales and detect conflicts. However, detailed entity interactions across all scales would introduce a large amount of redundant information and increase computational complexity. Many minor conflicts are already resolved by the \emph{multi-scale fusion reasoning} itself. Therefore, we propose a \emph{conflict detection} mechanism that provides secondary feedback based on the results of the \emph{multi-scale fusion reasoning} to address the conflicts arising from macro-micro variations at different scales. Specifically, we detect aligned entity pairs for each scale $\phi_\ell$ from the output of the \emph{multi-scale fusion reasoning} and check whether the same source entity aligns with different target entities across different scales. If such alignment exist, it is considered a conflict. Formally, given aligned entity pairs $\{\phi_\ell\}_{\ell=1}^3$ across all scales, we define the conflict detection function: 

\begin{equation}
\begin{aligned}
&\mathcal{D} = \{(e^s_i, e^t_j), (e^s_i, e^t_k) \mid \exists \ell_1, \ell_2 \in \{1,2,3\}:\\& (e^s_i, e^t_j) \in \phi_{\ell_1}, (e^s_i, e^t_k) \in \phi_{\ell_2}, \ell_1 \neq \ell_2, e^t_j \neq e^t_k\}.
\end{aligned}
\end{equation}

The mechanism groups conflicting entity pairs by source entity:

\begin{equation}
\begin{aligned}
C(e^s_i) = \{(e^s_i, e^t_j) \mid (e^s_i, e^t_j) \in \mathcal{D}\}.
\end{aligned}
\end{equation}

These conflicting sets $\{C(e^s_i)\}_{i=1}^{|E^s|}$ are then fed back to the \emph{multi-scale fusion reasoning} for resolution.

\myparagraph{Iterative Update} The fusion reasoning entity pairs $\phi_{f} = \bigcap_{\ell=1}^{3} \phi_\ell \cup \phi_C = \{(e^s_i, e^t_j) \mid e^s_i \in E^s, e^t_j \in E^t\}$ obtained from \emph{multi-scale fusion reasoning} are incorporated into the seed entity pairs and fed back to the \emph{multi-granular temporal encoder} and \emph{meta structural encoder} for updates, initiating a new round of iteration. After $I_n$ iterations, the \emph{adaptive integration} is able to generate high-quality alignment results.


\begin{table*}[ht!]
	\centering
	\caption{Dataset statistics~\cite{simplehhea,chatea,TEA-GNN,sun20}. 
    ``\textit{\#Ent}'', ``\textit{\#Rel.}'', ``\textit{\#Facts}'', ``\textit{\#T.Facts}'': The number of entities, relations, quadruples and quadruples with valid time interval in KG1 (KG2), respectively. \colorbox{blue!5}{``\textit{Temp.}'', ``\textit{Multi-Granularity}''}: Indicates whether the dataset includes temporal knowledge information and the dataset includes multi-granularity temporal knowledge information, respectively. ``\textit{\#Overlapping}'': Represents the proportion of overlapping temporal entities in KG1 and KG2. \colorbox{blue!15}{``\textit{Inter. Consis.}''}: Represents the proportion of aligned entities with consistent temporal intervals among all aligned entities.
    \colorbox{red!5}{``\textit{Multi-Source}'', ``\textit{MTF.\%}''}: Refers to whether both TKGs in the dataset are temporal incompleteness, and the average proportion of valid temporal facts in the two TKGs, respectively. 
    \colorbox{red!15}{``\textit{$\Delta$ T.F.\%}'', ``\textit{$\Delta$ T.D.\%}''}: Relative difference in valid temporal facts/density values between two KGs, using the KG with the smaller valid temporal facts/lower valid temporal density as the base. }
	\label{tab:stats}
	\adjustbox{max width=\textwidth}{
	\begin{tabular}{cc ccbbccnccooici}
		\toprule
		\rowcolor{white}\multicolumn{2}{c}{\bfseries Dataset} & \bfseries \#Ent. & \bfseries \#Rel. & \bfseries Temp. & \bfseries Multi-Granularity & \bfseries \#Seed & \bfseries \#Overlapping & \bfseries Inter. Consis. \downgood & \bfseries \#Facts & \bfseries \#T.Facts & \bfseries Multi-Source & \bfseries MTF.\% \downgood & \bfseries $\Delta$ T.F.\% \upgood & \bfseries \#T.Density & \bfseries $\Delta$ T.D.\% \upgood \\
		\cmidrule(lr){1-2} \cmidrule(lr){3-16}

        \multirow{2}{*}{\textbf{\enfr}} & EN & 15,000 & 193 & \xmark & \xmark & \multirow{2}{*}{15,000} & 100$\%$ &  & 96,318 & 0 &  &  &  & \xmark &  \\
                                       & FR & 15,000 & 166 & \xmark & \xmark &  & 100$\%$ & \multirow{-2}{*}{\xmark} & 80,112 & 0 & \multirow{-2}{*}{\xmark} & \multirow{-2}{*}{\xmark} & \multirow{-2}{*}{\xmark} & \xmark & \multirow{-2}{*}{\xmark} \\
		\midrule

        \multirow{2}{*}{\textbf{\dbpwiki}} & DBP  & 100,000 & 413 & \xmark & \xmark & \multirow{2}{*}{100,000} & 100$\%$ &  & 293,990 & 0 &  &  &  & \xmark &  \\
                                          & WIKI & 100,000 & 261 & \xmark & \xmark &  & 100$\%$ & \multirow{-2}{*}{\xmark} & 251,708 & 0 & \multirow{-2}{*}{\xmark} & \multirow{-2}{*}{\xmark} & \multirow{-2}{*}{\xmark} & \xmark & \multirow{-2}{*}{\xmark} \\
        \midrule

		\multirow{2}{*}{\textbf{\icewswiki}} & ICEWS & 11,047 & 272 & \cmark & \xmark & \multirow{2}{*}{5,058} & 45.79$\%$ & & 3,527,881 & 3,527,881 &  &  &  & 319.352 &  \\
                                           & WIKI  & 15,896 & 226 & \cmark & \xmark &  & 31.82$\%$ & \multirow{-2}{*}{55.63\%} & 198,257 & 51,002 & \multirow{-2}{*}{\xmark} & \multirow{-2}{*}{96.05\%} & \multirow{-2}{*}{6,817.1\%} & 3.208 & \multirow{-2}{*}{9,853.4\%} \\
        \midrule

		\multirow{2}{*}{\textbf{\icewsyago}} & ICEWS & 26,863 & 272 & \cmark & \xmark & \multirow{2}{*}{18,824} & 70.07$\%$ &  & 4,192,555 & 4,192,555 &  &  &  & 156.072 &  \\
                                           & YAGO  & 22,734 & 41  & \cmark & \xmark &  & 82.80$\%$ & \multirow{-2}{*}{7.39\%} & 107,118 & 30,240 & \multirow{-2}{*}{\xmark} & \multirow{-2}{*}{98.21\%}  & \multirow{-2}{*}{13,764.3\%} & 1.330 & \multirow{-2}{*}{11,633.3\%} \\
        \midrule

		\multirow{2}{*}{\textbf{\dicews}} & ICEWS & 9,517 & 247 & \cmark & \xmark & \multirow{2}{*}{8,566} & 90.01$\%$ &  & 307,552 & 307,552 &  &  &  & 32.316 &  \\
                                        & ICEWS & 9,537 & 246 & \cmark & \xmark &  & 89.82$\%$ & \multirow{-2}{*}{95.09\%} & 307,553 & 307,553 & \multirow{-2}{*}{\xmark} & \multirow{-2}{*}{100\%} & \multirow{-2}{*}{0\%} & 32.248 & \multirow{-2}{*}{0.2\%}\\
        \midrule

		\multirow{2}{*}{\textbf{\yw}} & YAGO & 49,629 & 11 & \cmark & \xmark & \multirow{2}{*}{49,172} & 99.08$\%$ &  & 221,050 & 221,050 &  &  &  & 4.454 &  \\
                                     & WIKI & 49,222 & 30 & \cmark & \xmark &  & 99.90$\%$ & \multirow{-2}{*}{93.63\%} & 317,814 & 317,814 & \multirow{-2}{*}{\xmark} & \multirow{-2}{*}{100\%} & \multirow{-2}{*}{43.8\%} & 6.457 & \multirow{-2}{*}{45.0\%} \\
        \midrule

		\multirow{2}{*}{\textbf{\Beta}} & WIKI & 42,666 & 257 & \cmark & \cmark & \multirow{2}{*}{40,364} & 94.60$\%$ & & 199,879 & 104,774 &  & &  & 2.456 &  \\
                                       & YAGO & 42,297 & 45  & \cmark & \cmark &  & 95.43$\%$ & \multirow{-2}{*}{55.12\%} & 162,320 & 69,896 & \multirow{-2}{*}{\cmark} & \multirow{-2}{*}{48.22\%} & \multirow{-2}{*}{49.9\%} & 1.653 & \multirow{-2}{*}{48.6\%} \\
        \midrule

		\multirow{2}{*}{\textbf{\wild}} & WIKI & 27,519 & 301 & \cmark & \cmark & \multirow{2}{*}{17,124} & 62.23$\%$ &  & 527,977 & 142,145 &  &  &  & 5.165 &  \\
                                       & YAGO & 26,975 & 40  & \cmark & \cmark &  & 63.48$\%$ & \multirow{-2}{*}{\textbf{5.27\%}} & 36,283 & 1,008 & \multirow{-2}{*}{\cmark} & \multirow{-2}{*}{\textbf{25.37\%}} & \multirow{-2}{*}{\textbf{14,001.7\%}} & 0.037 & \multirow{-2}{*}{\textbf{13,722.5\%}} \\
        \bottomrule
	\end{tabular}
	}
    \vspace{-9pt}
\end{table*}




\section{Dataset}\label{sec:dataset}

Currently, there is no standard benchmark that captures these real-world challenges and effectively evaluates existing methods. In this section, we propose two TKGA-Wild benchmarks and introduce the construction of datasets that better mirrors the challenges of TKGA-Wild.

\subsection{Dataset Construction} 
\label{sec:const}
\myparagraph{Motivation}
The most widely used TKGA datasets are \dicews and \yw.
The two TKG subsets in \dicews are derived from the same knowledge base, and thus there is merely minor structural or semantic heterogeneity between the two TKGs. 
In \yw, most entities are associated with the same temporal relation, e.g., the quadruples containing \texttt{playsFor} relation account for over 98\% of all the temporal quadruples in \yw. 
Besides, for the timestamps, they only retain the year information, while abandoning the fine-grained time information. These characteristics render the dataset overly simplistic and incapable of reflecting realistic TKGA scenarios.
In addition, recently proposed TKGA datasets such as \icewswiki and \icewsyago also fall short of reflecting realistic TKGA conditions due to their simple timestamps and overly complete multi-source temporal information. A case in point can be found in Fig.~\ref{fig:example}.

Hence, we aim to establish two news datasets to mirror the challenges of \emph{multi-scale temporal element entanglement (i.e., multi-granular temporal coexistence and temporal interval topological disparity)} and \emph{more realistic cross-source temporal structural imbalance (i.e., multi-source temporal structural incompleteness and temporal event density imbalance)} in TKGA in the wild. 

We illustrate the detailed construction processes of the datasets. For ease of explanation, we will take \Beta as an example in the following discussion, while the process for \wild is similar, with differences such as sampling the neighbors of entity pairs from the original KGs without enforcing the 1-to-1 assumption.

\myparagraph{Data Sources}
Our datasets are built upon Wikidata~\cite{DBLP:conf/iclr/LacroixOU20} and YAGO 4.5~\cite{YAGO45}. 
Specifically, in YAGO 4.5, facts and temporal facts are stored in separate files, and the start and end times of facts are independent. 
Thus, we integrate them to create the set of quadruples.

\myparagraph{Preliminary Seeds Generation}
We first select entities that are associated with multi-temporal relations from both KGs.
We retain those existing in the prior mappings~\footnote{\url{https://yago-knowledge.org/data/yago4.5/}} of YAGO entities to Wikidata QIDs, resulting in 26,594 entity pairs, which are termed as preliminary seeds.

\myparagraph{KG Subsets Generation}
Then, we extract quadruples associated with these preliminary seeds from respective KGs, leading to 131,853 quadruples from Wikidata and 89,702 quadruples from YAGO 4.5. 
We screen through the quadruples by selecting the entities that exist in the mapping, resulting in 40,364 pairs, which are used to retrieve the facts from respective KGs to form the final KG subset in each side, where Wikidata contains 199,879 quadruples and YAGO 4.5 contains 162,320 quadruples. 
Notably, in the final seed entity alignment pairs, 4,235 entities are unmatchable, indicating that there are no corresponding entities on the other side of the KG.
The specific dataset statistics are presented in Tab.~\ref{tab:stats}.

\begin{figure*}[t]
	\centering
	\subfigure[New alignment scenarios: Multi-to-multi.\label{eg1}]{
		\includegraphics[width=0.3\textwidth]{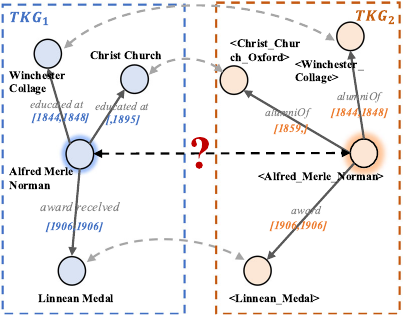}}
	~\quad
	\subfigure[New alignment scenarios: Multi-to-one.\label{eg2}]{
		\includegraphics[width=0.3\textwidth]{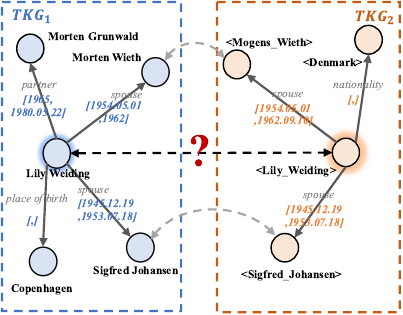}}
	~\quad
	\subfigure[New alignment scenarios: Multi-to-none.\label{eg3}]{
		\includegraphics[width=0.305\textwidth]{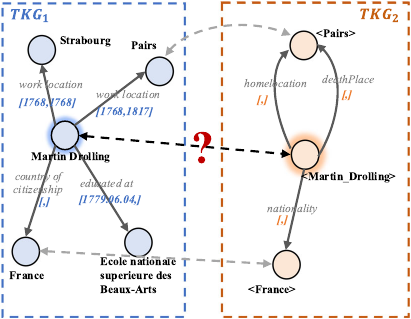}}
	\\
	\subfigure[New alignment scenarios: One-to-none.\label{eg4}]{
		\includegraphics[width=0.3\textwidth]{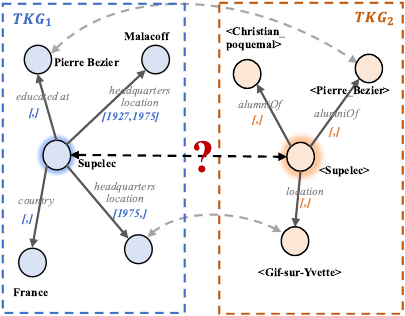}}
	~\quad\quad\quad\quad
	\subfigure[New alignment scenarios: None-to-none.\label{eg5}]{
		\includegraphics[width=0.3\textwidth]{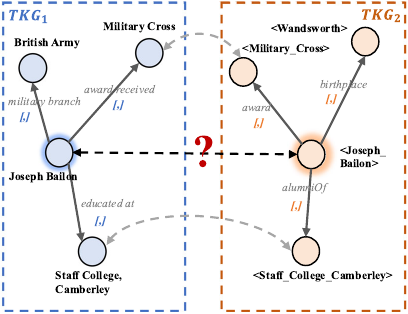}}
	\caption{New alignment scenarios in \Beta.}
	\label{fig:dis}
\end{figure*}
 
\subsection{Dataset Analysis}

Next, we conduct an in-depth analysis of \Beta and \wild, focusing on following key characteristics that reflect real-world TKGA challenges.

\myparagraph{Multi-Scale Temporal Element Distribution} \emph{\textbf{1) Multi-Granular Temporal Coexistence.}}
As shown in Tab.~\ref{tab:stats}, in the existing TKGA datasets, the timestamps are uniform and simplified, retaining only a single granularity of information. In contrast, both \Beta and \wild contain temporal information at multiple granularities. For example, the timestamps are simplified and only retain the single year information in  \yw dataset. In \wild, 83\% and 15\% of the temporal facts in the Wikidata and YAGO subsets, respectively, include month-date information. This diverse temporal representation better reflects real-world temporal patterns, where entities simultaneously engage in events with both coarse-grained and fine-grained temporal contexts. \emph{\textbf{2) Temporal Interval Topological Disparity.}} Furthermore, the topological relationships between temporal intervals in \Beta and \wild are highly diverse. As reported in Tab.~\ref{tab:stats}, only 5.27\% of the entity pairs in \wild share consistent temporal intervals across both TKGs. Additionally, due to its non 1-to-1 alignment setting, the entity overlap ratio across graphs is not ideally close to 100\%,, reinforcing the misalignment of temporal scopes and boundaries. More detailed formulations of the metrics can be found in technical report.


\myparagraph{Cross-Source Temporal Structural Imbalance Phenomenon} \emph{\textbf{1) Multi-Source Temporal Incompleteness.}} As shown in Tab.~\ref{tab:stats}, both \Beta and \wild exhibit significant temporal incompleteness across sources. The rate of multi-source temporal completeness reaches 48.22\% in \Beta and 25.37\% in \wild, far lower than the completeness levels of existing TKGA datasets. This incompleteness exacerbates the difficulty of learning alignment patterns (c.f. Section~\ref{sect:3.3}), as corresponding entities may have partially missing or non-overlapping temporal evidence. \emph{\textbf{2) Temporal Event Density Imbalance.}} Unlike conventional datasets where both valid temporal fact volume and density are balanced across KGs. For example, \dicews exhibits only 0\% and 0.2\% difference in valid temporal fact count and density, respectively. In contrast, \wild exhibits extreme variations in valid temporal fact change rates and entity density fluctuations. Valid temporal fact change rates and density fluctuations can reach up to 14,001.7\%, introducing a new level of alignment difficulty under asymmetric temporal dynamics.

\subsection{New Alignment Scenarios}\label{sect:3.3}
For existing datasets, the TKGA algorithms merely need to align the football players and teams between two KGs, as the majority of relations are \textit{member\_of\_sports\_team} and \textit{plays\_for}. 
While in \Beta and \wild, due to the introduction of more realistic quadruples, in addition
to the current alignment scenario where the entities to be aligned are only associated with one specific temporal relation, there are new temporal alignment scenarios emerging:

\textbf{Multi-to-multi/multi-to-one alignment}, where at least one of the entity to be aligned is associated with multiple temporal relations. 
In both cases, entities are involved with multiple temporal quadruples, and it is important to mine useful signals from the complex structure and reduce the noise brought by the heterogeneity of the relational and temporal information.  

\textbf{Multi-to-none/one-to-none alignment}, where at least one of the entity to be aligned is not associated with any temporal relations. In these cases, the key is to properly tackle the discrepancy of the temporal information and sufficiently exploit features modeled from other aspects. 

\textbf{None-to-none alignment}, where both entities are not associated with any temporal relation. In this case, temporal entity alignment is reduced to the general entity alignment task.

We present specific cases of multi-to-multi, multi-to-one, multi-to-none, one-to-none, and none-to-none alignment regarding the temporal relations in Fig.~\ref{fig:dis}.

\begin{table*}[t!]
	\caption{Main experiment results on TKGA datasets. Best results are highlighted in \textit{bold}, while runner-up results are \textit{underlined}. ``\textit{Struc.}, \textit{Temp.}, \textit{LLM.}, \textit{Retri.}'' indicate the use of structural information, temporal information, large language models, and a LLM-based retrieval-augmented generation process, respectively. These conventions apply to all subsequent tables.}
	\label{tb:main results}
	\centering
	\begin{adjustbox}{max width=\textwidth}
		\begin{tabular}{c|lp{0.4cm}p{0.4cm}p{0.4cm}p{0.4cm}cccccccccccc}
			\toprule
			\multicolumn{2}{c}{\multirow{2}{*}{\textbf{Models}}} &\multicolumn{4}{c}{\textbf{Settings}}&\multicolumn{3}{c}{\textbf{\wild}} &\multicolumn{3}{c}{\textbf{\Beta}} &\multicolumn{3}{c}{\textbf{\ywa(\ywas)}} &\multicolumn{3}{c}{\textbf{\da(\das)}}\cr
			\cmidrule(lr){3-6}\cmidrule(lr){7-9}\cmidrule(lr){10-12}\cmidrule(lr){13-15}\cmidrule(lr){16-18}
			\multicolumn{2}{c}{}&\textit{Struc.}&\textit{Temp.}&\textit{LLM.}&\textit{Retri.}&Hits@1 &Hits@10 &MRR &Hits@1 &Hits@10 &MRR &Hits@1 &Hits@10 &MRR &Hits@1 &Hits@10 &MRR\cr
			\midrule

			\multirow{2}{*}{\rotatebox{90}{Trans.}} &\MTransE&&&&&0.028&0.101&0.055&0.033&0.108&0.060&0.037&0.122&0.067&0.034&0.113&0.062\cr
			&\AlignE&\cmark&&&&0.108&0.319&0.252&0.229&0.424&0.296&0.552&0.704&0.605&0.170&0.381&0.242\cr
			
			\midrule
			\multirow{9}{*}{\rotatebox{90}{GNN}}&\GCNAlign&\cmark&&&&0.104&0.253&0.149&0.121&0.270&0.172&0.046&0.140&0.079&0.106&0.241&0.151\cr

		    &\mraea&\cmark&&&&0.301&0.492&0.385&0.407&0.611&0.478&0.642&0.816&0.703&0.466&0.716&0.550\cr 
            &\rrea&\cmark&&&&0.342&0.534&0.472&0.435&0.644&0.507&0.722&0.870&0.775&0.632&0.815&0.696\cr 
            &\DualAMN\textit{(basic)}&\cmark&&&&0.380&0.557&0.501&0.496&0.695&0.566&0.757&0.895&0.806&0.663&0.847&0.729\cr
            &\TEAGNN&\cmark&\cmark&&&0.426&0.613&0.541&0.557&0.709&0.611&0.723&0.870&0.775&0.806&0.901&0.843\cr 
            &\STEA&\cmark&\cmark&&&0.382&0.520&0.498&0.506&0.637&0.553&0.887&0.965&0.916&0.923&0.960&0.939\cr 
            &\dualmatch&\cmark&\cmark&&&0.474&0.661&0.530&0.650&0.765&0.690&0.945&0.984&0.960&0.952&0.974&0.961\cr 
            &\mgtea&\cmark&\cmark&&&0.556&0.698&0.646&0.686&0.795&0.724&0.947&0.982&0.960&0.951&0.974&0.960\cr 

            &\HTEA&\cmark&\cmark&&&0.532&0.612&0.568&0.661&0.742&0.695&0.932&0.963&0.956&0.928&0.955&0.941\cr
            
			\midrule
			\multirow{6}{*}{\rotatebox{90}{Other}}&\lightea&\cmark&\cmark&&&0.425&0.615&0.503&0.615&0.727&0.654&0.966&0.988&0.975&0.947&0.966&0.955\cr

            &\SimpleHHEA&&\cmark&&&0.608&0.681&0.645&0.762&0.835&0.797&0.156&0.368&0.209&0.849&0.947&0.895\cr
			&\SimpleHHEA\textit{(structure)}&\cmark&\cmark&&&0.591&0.674&0.621&0.730&0.808&0.751&0.121&0.303&0.170&0.836&0.938&0.874\cr

            &\nativerag$^{\textbf{$\star$}}$&\cmark&\cmark&\cmark&\cmark&0.531&-&-&0.588&-&-&0.102&-&-&0.652&-&-\cr

            &\self$^{\textbf{$\star$}}$&\cmark&\cmark&\cmark&&0.321&-&-&0.456&-&-&0.094&-&-&0.582&-&-\cr

            &\selfrag$^{\textbf{$\star$}}$&\cmark&\cmark&\cmark&\cmark&0.587&-&-&0.633&-&-&0.175&-&-&0.739&-&-\cr
            
            \midrule

			&\ourmodel (Ours) &\cmark&\cmark&\cmark&\cmark&\textbf{0.871}&\textbf{0.925}&\textbf{0.893}&\textbf{0.924}&\textbf{0.951}&\textbf{0.938}&\textbf{0.978}&\textbf{0.994}&\textbf{0.981}&\textbf{0.989}&\textbf{0.998}&\textbf{0.993}\cr

			\bottomrule
	\end{tabular}
	\end{adjustbox}
\end{table*}

\section{Experiments}
\label{sec:experiment}
This section begins by outlining the experimental setup in Section~\ref{sec5:setting}. Subsequently, Sections~\ref{sec5:main} to \ref{sec5:more} present a comprehensive empirical evaluation of the proposed datasets and framework from multiple perspectives, aiming to address the following research questions:
\squishlist

\item \textbf{RQ1: How effective is the overall TKGA-Wild framework? (Section \ref{sec5:main})}

\item \textbf{RQ2: How difficult and useful are the \Beta and \wild datasets? (Section \ref{sec5:difficulty})}

\item \textbf{RQ3: What is the effectiveness of each component within \ourmodel? (Section \ref{sec5:ablation})} 

\item \textbf{RQ4: Does the \ourmodel successfully balance alignment accuracy and efficiency in TKGA-Wild? (Section \ref{sec5:efficiency})}

\item \textbf{RQ5: To what degree is \ourmodel adaptable and robust across varying foundation models and more complex TKGA-Wild scenarios? (Section \ref{sec5:more})} 

\squishend

\subsection{Experimental Setting}\label{sec5:setting}
In our experiments, we conducted comprehensive evaluations on \textbf{eight datasets}, including \Beta, \wild, and six current datasets (as detailed in Tab.~\ref{tab:stats} and Section~\ref{sec:background}). We compared our approach against \textbf{24 SOTA and classic baseline methods}. By reviewing both TKGA and non-temporal KGA methods, we categorized the 24 representative methods. These include translation-based methods (i.e., ``Trans.'') such as \MTransE~\cite{MTransE}, \AlignE~\cite{BootEA} and \BootEA~\cite{BootEA}; GNN-based methods (i.e., ``GNN'') like \GCNAlign~\cite{GCN-Align}, \mraea~\cite{mraea}, \rrea~\cite{rrea}, \RDGCN~\cite{RDGCN}, \DualAMN~\cite{Dual-AMN}, \TEAGNN~\cite{TEA-GNN}, \TREA~\cite{TREA}, \STEA~\cite{STEA}, \Dualmatch\cite{dualmatch}, \MGTEA\cite{Zeng2024BenchmarkF}, \HTEA\cite{wsdm}; as well as other TKGA and non-temporal KGA methods (i.e., ``Other'') such as \lightea~\cite{lightea}, \BERT~\cite{bert}, \FuAlign~\cite{Fualign}, \BERTINT~\cite{BERT-INT}, \PARIS~\cite{PARIS,simplehhea}, \SimpleHHEA~\cite{simplehhea}, and \ChatEA~\cite{chatea}. Among them, \TEAGNN~\cite{TEA-GNN}, \TREA~\cite{TREA}, \STEA~\cite{STEA}, \Dualmatch~\cite{dualmatch}, \MGTEA~\cite{Zeng2024BenchmarkF}, \lightea~\cite{lightea}, \SimpleHHEA~\cite{simplehhea}, \ChatEA~\cite{chatea}, and \HTEA\cite{wsdm} are the advanced and representative TKGA methods.

Since there are currently no LLM-based retrieval-augmented generation (RAG) methods specifically designed for TKGA, we additionally include three general methods inspired by LLM or RAG paradigms to enhance comprehensiveness (with `$^{\textbf{$\star$}}$''):
1) \nativerag~\cite{ragsurvey2,naive}, a basic LLM-based RAG approach that first retrieves relevant information based on a user query and then generates answers using the retrieved content; and
2) \self~\cite{selfcot}, a chain-of-thought baseline that produces multiple reasoning paths and selects the most frequent answer as the final output; and
3) \selfrag~\cite{selfrag,ICLRRAG25}, a self-reflective RAG method aimed at improving the generation quality of LLMs.

Following previous works~\cite{Zeng2024BenchmarkF,simplehhea}, we use \textit{Hits@N} (where $N=1, 5, 10$) and \textit{Mean Reciprocal Rank} (MRR) as the fundamental evaluation metrics. All LLMs in Tab.~\ref{tb:main results} and Tab.~\ref{tb:main results2} are implemented with the consistent model version, GPT-4 (gpt-4-0125-preview). For subsequent experiments, unless otherwise stated, we use GPT-3.5 (gpt-3.5-turbo-1106) as our standard LLM configuration due to its lower computational cost.
More intricate settings can be found in technical report.

\subsection{Main Results}\label{sec5:main}
As shown in Tab.~\ref{tb:main results} and Tab.~\ref{tb:main results2}, to address \textbf{RQ1}, we have performed a comprehensive comparison, and the experiments show that \ourmodel significantly outperform the all baselines groups, especially in TKGA-Wild scenarios. These results lead to several key observations:

\myparagraph{Comparison with All Baseline Methods under the TKGA-Wild Scenarios}
We first compare our \ourmodel framework against all baseline methods on \wild and \Beta.

1) Compared with translation-based, GNN-based, and other TKGA or non-temporal KGA approaches under the TKGA-Wild scenarios, \ourmodel consistently achieves superior performance. Specifically, it yields improvements of 43.3\% and 21.3\% in Hits@1 over the best SOTA baseline (\SimpleHHEA) on \wild and \Beta, respectively. These results underscore \ourmodel's effectiveness in addressing key challenges in TKGA-Wild scenarios, including \emph{multi-scale temporal element entanglement (i.e., multi-granular temporal coexistence and temporal interval topological disparity)} and \emph{cross-source temporal structural imbalance (i.e., multi-source temporal incompleteness and temporal event density imbalance)}. This firmly establishes \ourmodel as the new SOTA in the TKGA-Wild scenarios.

2) In comparison to LLM-based RAG methods (e.g., \nativerag and \selfrag), we observe that all retrieval-based models consistently outperform approaches that rely solely on LLMs (\self). This finding underscores the critical importance of framing the TKGA-Wild task as a retrieval-centric problem to boost overall performance.

3) Compared to \nativerag and \selfrag(both integrating LLMs and retrieval mechanisms), \ourmodel demonstrates superior performance, achieving up to a 50.6\% improvement. This advantage stems from our architecture, specifically designed for the TKGA-Wild task from a hypergraph retrieval-augmented perspective. It incorporates six new modules: \emph{meta structural encoder}, \emph{multi-granular temporal encoder}, \emph{adaptive integration}, \emph{scale-adaptive entity projection}, \emph{multi-scale hypergraph retrieval}, and \emph{multi-scale interaction-augmented fusion}. These innovations enable a deeper understanding of the essential characteristics of TKGA in real-world settings, facilitating both efficient and accurate temporal alignment.

\begin{table*}[t!]
	\caption{More experiment results on different TKGA and KGA datasets.}
	\label{tb:main results2}
	\centering
	\begin{adjustbox}{max width=\textwidth}
		\begin{tabular}{c|lp{0.4cm}p{0.4cm}p{0.4cm}p{0.4cm}cccccccccccc}
			\toprule
			\multicolumn{2}{c}{\multirow{2}{*}{\textbf{Models}}} &\multicolumn{4}{c}{\textbf{Settings}}&\multicolumn{3}{c}{\textbf{\icewswiki}} &\multicolumn{3}{c}{\textbf{\icewsyago}}&\multicolumn{3}{c}{\textbf{\enfr}} &\multicolumn{3}{c}{\textbf{\dbpwiki}} \cr
			\cmidrule(lr){3-6}\cmidrule(lr){7-9}\cmidrule(lr){10-12}\cmidrule(lr){13-15}\cmidrule(lr){16-18}
			\multicolumn{2}{c}{}&\textit{Struc.}&\textit{Temp.}&\textit{LLM.}&\textit{Retri.}&Hits@1 &Hits@10 &MRR &Hits@1 &Hits@10 &MRR &Hits@1 &Hits@10 &MRR &Hits@1 &Hits@10 &MRR\cr

            \midrule
			\multirow{3}{*}{\rotatebox{90}{Trans.}} &\MTransE&&&&&0.021&0.158&0.068&0.012&0.084&0.040&0.247&0.577&0.360&0.281&0.520&0.363\cr
			&\AlignE&\cmark&&&&0.057&0.261&0.122&0.019&0.118&0.055&0.481&0.824&0.599&0.566&0.827&0.655\cr
			&\BootEA&\cmark&&&&0.072&0.275&0.139&0.020&0.120&0.056&0.653&0.874&0.731&0.748&0.898&0.801\cr
			\midrule
			\multirow{9}{*}{\rotatebox{90}{GNN}}&\GCNAlign&\cmark&&&&0.046&0.184&0.093&0.017&0.085&0.038&0.411&0.772&0.530&0.494&0.756&0.590\cr
			&\RDGCN&\cmark&&&&0.064&0.202&0.096&0.029&0.097&0.042&0.873&0.950&0.901&0.974&0.994&0.980\cr
			&\DualAMN\textit{(basic)}&\cmark&&&&0.077&0.285&0.143&0.032&0.147&0.069&0.756&0.948&0.827&0.786&0.952&0.848\cr
			&\DualAMN\textit{(semi.)}&\cmark&&&&0.037&0.188&0.087&0.020&0.093&0.045&0.840&0.965&0.888&0.869&0.969&0.908\cr
			&\DualAMN\textit{(name)}&\cmark&&&&0.083&0.281&0.145&0.031&0.144&0.068&0.954&0.994&0.970&0.983&0.996&0.991\cr
   		
            &\TEAGNN&\cmark&\cmark&&&0.063&0.253&0.126&0.025&0.135&0.064&-&-&-&-&-&-\cr 
		    &\TREA&\cmark&\cmark&&&0.081&0.302&0.155&0.033&0.150&0.072&-&-&-&-&-&-\cr 
            &\STEA&\cmark&\cmark&&&0.079&0.292&0.152&0.033&0.147&0.073&-&-&-&-&-&-\cr 
            &\Dualmatch&\cmark&\cmark&&&0.104&0.354&0.179&0.042&0.160&0.081&
            -&-&-&-&-&-\cr 
			\midrule
			\multirow{10}{*}{\rotatebox{90}{Other}}&\BERT&&&&&0.546&0.687&0.596&0.749&0.845&0.784&0.937&0.985&0.956&0.941&0.980&0.963\cr
			&\FuAlign&\cmark&&&&0.257&0.570&0.361&0.326&0.604&0.423&0.936&0.988&0.955&0.980&0.991&0.986\cr
			&\BERTINT&\cmark&&&&0.561&0.700&0.607&0.756&0.859&0.793&\underline{0.990}&\underline{0.997}&0.993&\textbf{0.996}&0.997&\underline{0.996}\cr
   		
            &\PARIS&\cmark&&&&0.672&-&-&0.687&-&-&0.902&-&-&0.963&-&-\cr
      
			&\SimpleHHEA&&\cmark&&&0.720&0.872&0.754&0.847&0.915&0.870&0.948&0.991&0.960&0.967&0.988&0.979\cr
			&\SimpleHHEA\textit{(structure)}&\cmark&\cmark&&&0.639&0.812&0.697&0.749&0.864&0.775&0.959&0.995&0.972&0.975&0.991&0.988\cr



            
            &\ChatEA&\cmark&\cmark&\cmark&&\underline{0.880}&\underline{0.945}&\underline{0.912}&\underline{0.935}&\underline{0.955}&\underline{0.944}&\underline{0.990}&\textbf{1.000}&\underline{0.995}&\underline{0.995}&\textbf{1.000}&\textbf{0.998}\cr

            &\nativerag$^{\textbf{$\star$}}$&\cmark&\cmark&\cmark&\cmark&0.723&-&-&0.736&-&-&0.894&-&-&0.912&-&-\cr

            &\self$^{\textbf{$\star$}}$&\cmark&\cmark&\cmark&&0.718&-&-&0.725&-&-&0.885&-&-&0.903&-&-\cr

            &\selfrag$^{\textbf{$\star$}}$&\cmark&\cmark&\cmark&\cmark&0.772&-&-&0.762&-&-&0.918&-&-&0.937&-&-\cr
            
            \midrule

			&\ourmodel (Ours) &\cmark&\cmark&\cmark&\cmark&\textbf{0.984}&\textbf{0.996}&\textbf{0.989}&\textbf{0.972}&\textbf{0.997}&\textbf{0.983}&\textbf{0.994}&\textbf{1.000}&\textbf{0.997}&\textbf{0.996}&\underline{0.999}&\textbf{0.998}\cr

			\bottomrule
	\end{tabular}
	\end{adjustbox}
\end{table*}







\myparagraph{Comparison with All Baselines in Representative TKGA Scenarios}
We further evaluate our approach against a wide range of advanced and representative baseline methods across four widely-used TKGA datasets, i.e., \ywa,\da, \icewswiki, and \icewsyago. As shown in Tab.~\ref{tb:main results} and Tab.~\ref{tb:main results2}, \ourmodel consistently outperforms the existing SOTA models on all four datasets, achieving up to a 11.8\% improvement in Hits@1. These results highlight the practical effectiveness and robustness of \ourmodel in various representative TKGA scenarios.

\myparagraph{Comparison with All Baselines in Non-Temporal KGA Scenarios}
To further validate the versatility of our model, we extend the evaluation to two classic non-temporal KGA datasets, \enfr and \dbpwiki. We compare \ourmodel with leading non-temporal KGA methods as well as TKGA models. As shown in Tab.~\ref{tb:main results2}, \ourmodel also achieves competitive or even superior performance in non-temporal settings, reaching 100\% Hits@10 on \enfr. These findings indicate that \ourmodel not only excels in both complex and simplified TKGA scenarios but is also highly effective and scalable in non-temporal KGA tasks.

\subsection{Analysis of Dataset Difficulty} \label{sec5:difficulty}

To address \textbf{RQ2}, we conduct an in-depth analysis of the \Beta and \wild datasets in terms of their alignment difficulty and practical utility. The goal is to understand how these datasets reflect the complexities of real-world TKGA scenarios and how they compare to existing benchmarks.

\myparagraph{Comparison with Existing TKGA Datasets} 
As shown in Tab.~\ref{tb:main results} and Tab.~\ref{tb:main results2}, we compare the performance of \ourmodel and other SOTA solutions on standard TKGA benchmarks, including \ywa, \da, \icewswiki, and \icewsyago. These datasets yield exceptionally high performance across all methods, with Hits@1 scores exceeding 97\% in most cases, suggesting they pose limited challenge to SOTA models. In contrast, both \wild and \Beta present substantially greater difficulty: the best baseline method achieves only around 60.8\% Hits@1 on \wild , indicating that these datasets better reflect the complexities of realistic TKGA settings. This highlights the need for more challenging benchmarks like TKGA-Wild to drive meaningful progress.

\myparagraph{Comparison with Non-Temporal KGA Datasets} 
To further contextualize the difficulty of \wild and \Beta, we evaluate both temporal and non-temporal KGA models on two classic non-temporal KGA datasets, \enfr and \dbpwiki. As reported in Tab.~\ref{tb:main results2}, both \ourmodel and the current best-performing model \ChatEA attain near-perfect results (over 99\% Hits@1), underscoring the relative simplicity of these static KGA tasks. This sharp contrast reinforces the value of TKGA-Wild datasets as challenging and realistic benchmarks that push the limits of current approaches in temporal alignment, particularly under conditions of \emph{multi-scale temporal element entanglement (i.e., multi-granular temporal coexistence and temporal interval topological disparity)} and \emph{cross-source temporal structural imbalance (i.e., multi-source temporal incompleteness and temporal event density imbalance)}.

\begin{table}[t]
\centering
\caption{The results of ablation study.}
\vspace{-4pt}
\setlength{\tabcolsep}{3pt}
\resizebox{1\linewidth}{!}{
\begin{tabular}{l|cccc}
    \toprule
    \multirow{2}{*}{\ \ \ \ \ \runhao{\textbf{Settings}}} &\multicolumn{4}{c}{\runhao{\wild}} \cr
    \cmidrule(lr){2-5}
    & {\runhao{Hits@1}} & \runhao{Hits@5} & \runhao{Hits@10} & \runhao{MRR}\cr
    \midrule

    \multicolumn{5}{c}{\cellcolor{lightgray!50}\textbf{\textit{\runhao{Group 1}}}} \cr
    
    \midrule
    \ourmodel & \textbf{0.859}&\textbf{0.886} & \textbf{0.904}&\textbf{0.878} \cr  
    - \textit{w/o} \emph{Multi-Granular Temporal Encoder}& \runhao{0.812} & \runhao{0.847} & \runhao{0.866} & \runhao{0.841}\cr
    \runhao{- \textit{w/o} Year Granularity}& \runhao{0.821} & \runhao{0.863} & \runhao{0.875} & \runhao{0.852}\cr
    \runhao{- \textit{w/o} Date Granularity}& \runhao{0.847} & \runhao{0.868} & \runhao{0.884} & \runhao{0.859}\cr

    \midrule

    \multicolumn{5}{c}{\cellcolor{lightgray!50}\textbf{\textit{\runhao{Group 2}}}} \cr
    
    \midrule
    
    \ourmodel & \textbf{0.859}&\textbf{0.886} & \textbf{0.904}&\textbf{0.878} \cr  
    
    - \textit{w/o} \emph{Scale-Adaptive Entity Projection}& 0.803 & 0.839 & 0.847 & 0.822\cr

    - \textit{w/o} Adaptive Time Projection& 0.831 & 0.863 & 0.874 & 0.859\cr

    - \textit{w/o} Adaptive Relation Projection& 0.848 & 0.871 & 0.886 & 0.869\cr

    \midrule

    \multicolumn{5}{c}{\cellcolor{lightgray!50}\textbf{\textit{\runhao{Group 3}}}} \cr
    
    \midrule
    \ourmodel & \textbf{0.859}&\textbf{0.886} & \textbf{0.904}&\textbf{0.878} \cr    
    - \textit{w/o} \emph{Multi-Scale Hypergraph Retrieval}& 0.764 & 0.807 & 0.830 & 0.796\cr

    - \textit{w/o} Multi-Scale Hypergraph& 0.791 & 0.842 & 0.859 & 0.838\cr

    \midrule

    \multicolumn{5}{c}{\cellcolor{lightgray!50}\textbf{\textit{Group 4}}} \cr
    
    \midrule
    \ourmodel & \textbf{0.859}&\textbf{0.886} & \textbf{0.904}&\textbf{0.878} \cr    
    - \textit{w/o} \emph{Multi-Scale Interaction-Augmented Fusion}& 0.611 & 0.664 & 0.698 & 0.645\cr

    - \textit{w/o} Intra-Scale Interaction& 0.827 & 0.852 & 0.864 & 0.849\cr
    
    - \textit{w/o} Multi-Scale Fusion Reasoning& 0.628 & 0.685 & 0.712 & 0.653\cr

    - \textit{w/o} Conflict Detection& 0.822 & 0.859 & 0.873 & 0.856\cr

    \bottomrule
\end{tabular}}
\label{tb:ablation}
\end{table}

\subsection{Ablation Study} \label{sec5:ablation}
To address \textbf{RQ3}, we conduct an ablation study organized into four groups based on the architecture of the proposed model. The experimental results, summarized in Tab.~\ref{tb:ablation}, yield the following key insights:

\myparagraph{Group 1} Removing the \emph{multi‑granular temporal encoder} module (\ourmodel \textit{w/o} \emph{Multi-Granular Temporal Encoder}) entirely leads to a 5.5\% relative drop in Hits@1, indicating that explicitly modeling multi-granular temporal information is crucial for improving reasoning accuracy. Further, separately ablating the year granularity and date granularity encoders shows that both granularities contribute positively, with the year granularity encoding having the most significant impact. This is because the year granularity can mitigate the issues caused by inconsistencies in temporal granularity and time label values. Besides, the dataset usually contains richer time information at the year granularity.

\myparagraph{Group 2} Disabling the \emph{scale‑adaptive entity projection} module (\ourmodel \textit{w/o} \emph{Scale‑Adaptive Entity Projection}) results in a 6.5\% decrease in Hits@1, validating the effectiveness of dynamically mapping entities to latent semantic spaces based on scale in addressing disparities in \emph{complex temporal interval topological disparity} and \emph{temporal event density distributions}. Further removal of its submodules (adaptive time projection and adaptive relation projection), leads to performance drops of 3.3\% and 1.3\%, respectively, highlighting the complementary roles of the temporal and relational adaptive pathways in supporting scale-aware decomposition and representation learning.

\myparagraph{Group 3} Eliminating the \emph{multi‑scale hypergraph retrieval} module (\ourmodel \textit{w/o} \emph{Multi‑Scale Hypergraph Retrieval}) and its components causes a substantial drop of up to 11\% in Hits@1. This underscores the indispensable role of high-order hyperedges in providing rich, interconnected semantic cues for temporal reasoning. The retrieval mechanism plays a critical role in injecting these cues effectively into the multi-scale hypergraph structure and downstream inference processes, thereby better capturing \emph{complex temporal interval topological disparity} and balancing \emph{temporal event density distributions}.

\myparagraph{Group 4} Removing the \emph{multi-scale interaction-augmented fusion} module (\ourmodel \textit{w/o} \emph{Multi-Scale Interaction-Augmented Fusion}) and its internal components results in a performance degradation ranging from 3.7\% to 28.9\%, highlighting the decisive role this module plays in aligning and reconciling cross-scale information. It also effectively alleviates the learning difficulties associated with \emph{multi-source temporal incompleteness-induced alignment pattern learning} and \emph{temporal event density imbalance}.

In summary, \ourmodel leverages a four-layer complementary design comprising multi-granular temporal encoding, scale-adaptive projection, hypergraph retrieval, and interaction-augmented fusion. Each module plays a critical role in enhancing the model’s understanding of the TKGA-Wild setting. Any further simplification compromises the synergistic mechanisms at the core of \ourmodel, making it inadequate for the complex modeling requirements of TKGA-Wild.

\begin{figure}[t]
    \centering
    \includegraphics[width=0.95\columnwidth]{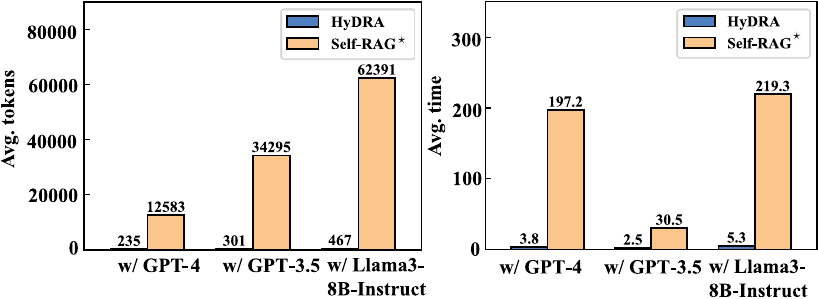}
    \vspace{-3pt}
    \caption{Efficiency analysis of \ourmodel and \selfrag$^{\textbf{$\star$}}$ on \wild. \emph{Avg.tokens}: the average tokens cost of the model per target entity. \emph{Avg.time}: the average time (seconds) cost of the model per target entity.}
    \label{fig:efficiency_compared}
    \vspace{-5pt}
\end{figure}

\begin{table}[t]
\centering
\caption{Accuracy and efficiency analysis of \ourmodel and baselines on \wild.}
\vspace{-6pt}
\setlength{\tabcolsep}{3pt}
\resizebox{0.9\linewidth}{!}{
\begin{tabular}{l|ccc|cc}
    \toprule
    \runhao{\multirow{2}{*}{\ \ \ \ \ \textbf{Settings}}} &\multicolumn{3}{c}{\runhao{\wild (Accuracy)}} &\multicolumn{2}{|c}{\runhao{Cost (Efficiency)}}\cr
    \cmidrule(lr){2-4}\cmidrule(lr){5-6}
    & \runhao{Hits@1} & Hits@10 & MRR & Avg. tokens & Avg. time (seconds)\cr
    \midrule
        
    \SimpleHHEA\textit{(structure)} & 0.591& 0.674& 0.621 & -& \textbf{1.3}\cr   

    \nativerag$^{\textbf{$\star$}}$ & 0.531& -& - & 1,782& 10.2\cr

    \self$^{\textbf{$\star$}}$ & 0.321& -& - & 13,488&20.7\cr   
    
    \selfrag$^{\textbf{$\star$}}$ & 0.587&-&- & 34,295& 30.5\cr   

    \midrule
    \ourmodel (Ours) & \textbf{0.859}& \textbf{0.904}& \textbf{0.878} & \textbf{301}& 2.5\cr
    \bottomrule
\end{tabular}}
\label{tb: compare_cost}
\vspace{-7pt}
\end{table}

\subsection{Efficiency Analysis}\label{sec5:efficiency}
To comprehensively address \textbf{RQ4}, we extended our evaluation beyond performance metrics by systematically analyzing the computational efficiency and token consumption of the proposed method.

As illustrated in Fig.~\ref{fig:efficiency_compared} and Tab.~\ref{tb: compare_cost}, our method demonstrates remarkable efficiency while attaining state-of-the-art accuracy. In comparison to \selfrag (utilizing GPT-3.5), our approach reduces execution time by 85.0\% (2.5 s vs. 30.5 s) and token consumption by 99.1\% (301 vs. 34,295), all while achieving a 46.3\% relative improvement in Hits@1. Although it exhibits marginally higher latency than the lightweight \SimpleHHEA model, our framework surpasses it with a substantial 70.8\% relative gain in Hits@1 accuracy. These results underscore the effectiveness of our approach in overcoming performance bottlenecks on the TKGA-Wild benchmark.

\begin{table}[t]
\centering
\caption{\ourmodel's performance with different LLM configurations. ``- \textit{w/}'' refers to the LLM's base model selected in \ourmodel.}
\setlength{\tabcolsep}{3pt}
\resizebox{0.9\linewidth}{!}{
\begin{tabular}{l|ccc|ccc}
    \toprule
    \multirow{2}{*}{\ \ \ \ \ \textbf{Settings}} &\multicolumn{3}{c}{\wild} &\multicolumn{3}{|c}{\icewswiki}\cr
    \cmidrule(lr){2-4}\cmidrule(lr){5-7}
    & {Hits@1} & {Hits@10} & {MRR} & {Hits@1} & {Hits@10} & {MRR}\cr
    \midrule
    \ourmodel & & & & \cr    
    - \textit{w/} \texttt{LLMs}: GPT-4\tablefootnote{gpt-4-0125-preview from OpenAI API, \url{ https://openai.com/api/}} & \underline{0.871} & \underline{0.925} & \underline{0.893} & \underline{0.984} & \underline{0.996} & \underline{0.989}\cr
    - \textit{w/} \texttt{LLMs}: GPT-3.5\tablefootnote{gpt-3.5-turbo-0125 from OpenAI API, \url{https://openai.com/api/}} & 0.859 & 0.904 & 0.878 & 0.978 & 0.989 & 0.981\cr

    - \textit{w/} \texttt{LLMs}: Llama3-8B-Instruct\tablefootnote{\url{ https://huggingface.co/meta-llama/Meta-Llama-3-8B-Instruct}} & 0.639 & 0.776 & 0.653 & 0.734 & 0.875 & 0.807\cr
    - \textit{w/} \texttt{LLMs}: Claude 3.5 Sonnet\tablefootnote{claude-3-5-sonnet-20240620 from Claude API, \url{ https://www.anthropic.com/claude/sonnet}} & \textbf{0.887} & \textbf{0.938} & \textbf{0.912} & \textbf{0.987} & \textbf{0.998} & \textbf{0.991} \cr
    \bottomrule
\end{tabular}}
\label{tb: llm_Generality}
\vspace{-10pt}
\end{table}

\begin{figure*}[t]
    \centering
    \includegraphics[width=1\textwidth]{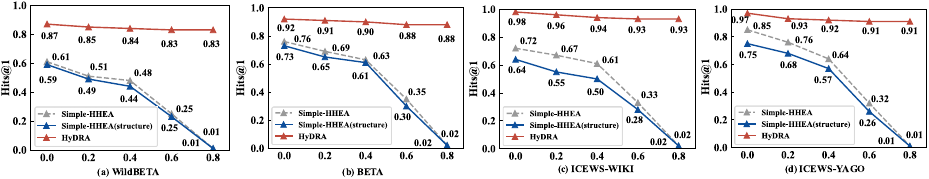}
    \vspace{-5pt}
    \caption{Adaptability comparison of \ourmodel and two representative SOTA embedding methods under varying noisy embeddings conditions on \wild, \Beta, \icewswiki, and \icewsyago.}
    \label{fig:compared_name}
    \vspace{-10pt}
\end{figure*}

\subsection{Generalization and Adaptability Analysis}\label{sec5:more}
To address \textbf{RQ5}, we conduct a series of experiments designed to evaluate the generalization capability and adaptability of \ourmodel in diverse and challenging TKGA-Wild scenarios.

\myparagraph{Model Generality under Varying LLM Configurations}\label{sec5:generality} 
To assess the adaptability and generalization capability of our proposed framework, we experiment with a range of base LLMs within the \emph{multi-scale interaction-augmented fusion} component. As illustrated in Tab.~\ref{tb: llm_Generality}, \ourmodel achieves state-of-the-art results when powered by Claude 3.5 Sonnet as the underlying LLM agent. Interestingly, performance improvements appear to align with increases in both model scale and version updates (e.g., GPT-3.5 vs. GPT-4), which is in line with previous findings~\cite{chatea}. This trend indicates that \ourmodel is well-positioned to benefit from ongoing advancements in LLM technology, affirming its strong potential for scalability and robust generalization in TKGA-Wild scenarios.

\myparagraph{Adaptability of \ourmodel under Varying Noisy Inputs}
To comprehensively evaluate the adaptability of \ourmodel in complex and noise-perturbed TKGA-Wild scenarios, we designed a series of experiments simulating the degradation of entity embeddings. Specifically, we injected uniformly distributed random noise with varying ratios (from 0\% to 80\%) into the initial entity embeddings learned by two representative SOTA embedding methods (\SimpleHHEA and \SimpleHHEA\textit{(structure)}) and \ourmodel. This setup aims to simulate highly complex alignment environments characterized by \emph{multi-scale temporal element entanglement (i.e., multi-granular temporal coexistence and temporal interval topological disparity)} and \emph{cross-source temporal structural imbalance (i.e., multi-source temporal incompleteness and temporal event density imbalance)} under enhanced noise conditions.

As shown in Fig.~\ref{fig:compared_name}, the performance of \SimpleHHEA degrades significantly as the noise level increases, revealing its vulnerability in handling highly complex temporal alignment scenarios. In contrast, \ourmodel consistently exhibits remarkable resilience. This robustness is attributed to its \emph{multi-scale hypergraph retrieval-augmented} paradigm and multi-scale hypergraph reasoning architecture, which enable adaptive representation learning in multi-scale temporal contexts. By dynamically retrieving and learning multi-scale hypergraphs and incorporating a multi-scale interaction-augmented fusion mechanism, \ourmodel effectively captures high-order temporal dependencies and substantially mitigates the negative effects caused by embedding quality degradation.

This experiment demonstrates the strong adaptability of \ourmodel when confronted with noisy, incomplete, and heterogeneous data, further validating its practical potential for deployment in real-world TKGA-Wild scenarios.

\myparagraph{Performance Analysis under More Complex New Alignment Scenarios}
Tab.~\ref{tab:scenarios} presents a detailed comparison of performance across six newly proposed, complex TKGA scenarios (refer to Section~\ref{sect:3.3} for more details). It is evident that \ourmodel consistently achieves superior performance across all alignment settings.

In particular, the baseline results show that the Hits@1 metric exceeds 90\% in the multi-to-multi, multi-to-one, and one-to-one alignment scenarios. This high performance can largely be attributed to the availability of rich temporal information. In these cases, each entity is associated with at least one temporal relation, thereby enhancing the alignment accuracy by leveraging temporal context.

In contrast, performance substantially declines in the multi-to-none, one-to-none, and none-to-none scenarios, with Hits@1 dropping below 30\%. This degradation is primarily due to the strong reliance of existing SOTA TKGA models on temporal information. These models struggle in the absence of Multi-Source Temporal Incompleteness and Complex Scale-Density Temporal Distribution, which increases the heterogeneity of temporal features among entities and impairs the alignment process.

Remarkably, \ourmodel demonstrates a significant performance gain of up to 286.0\% in Hits@1, highlighting its robustness and adaptability in addressing the challenges posed by these complex and underexplored TKGA-Wild scenarios.

\begin{table}
\centering
\caption{The Hits@1 results under different scenarios in \Beta .}
\begin{tabular}{ccccc}
\toprule 
               & LightTEA & DualMatch & MGTEA & \ourmodel  \\
\midrule 
Multi-to-multi   & 86.11   & 88.50    & 96.30  & \textbf{96.58} \\
Multi-to-one  & 82.79   & 85.37    & 89.19  & \textbf{95.34}  \\
Multi-to-none    & 17.24   & 24.23    & 27.04 & \textbf{83.89} \\
One-to-one & 87.84   & 89.98    & 91.76  & \textbf{96.58}\\
One-to-none   & 11.68   & 16.94    & 21.02  & \textbf{81.13} \\
None-to-none     & 16.63   & 22.06    & 26.46 & \textbf{83.56} \\
\bottomrule 
\end{tabular}
\label{tab:scenarios}%
\end{table}


\section{Conclusion and Future Work}\label{sec:con}
To the best of our knowledge, we are among the first to analyze the challenges of TKGA-Wild and attempt to address them, establishing a foundational approach and benchmark for future explorations in this more realistic area. To address the challenges of \emph{multi-scale temporal element entanglement (i.e., multi-granular temporal coexistence and temporal interval topological disparity)} and \emph{cross-source temporal structural imbalance (i.e., multi-source temporal incompleteness and temporal event density imbalance)} in TKGA-Wild,  we offer a new and effective TKGA-Wild solution, \ourmodel, the first to reformulate the task via \emph{multi-scale hypergraph retrieval-augmented generation} to address the challenges of TKGA-Wild.  \ourmodel effectively models complex structural dependencies and multi-granular temporal features, while also resolving temporal topological disparities and balancing event density distributions, thereby enhancing temporal alignment performance under wild conditions.
We further design a \emph{scale-weave synergy} mechanism within \ourmodel, which enables both \emph{intra-scale interactions} and cross-scale \emph{conflict resolution}. This mechanism mitigates the adverse effects of incomplete and fragmented temporal information from multiple sources and enhances the model's robustness to inconsistencies arising from dense and irregular temporal event distributions.

Currently, there is no standard benchmark that captures these challenges of TKGA-Wild and effectively evaluates existing methods. To this end, we establishes two new TKGA-Wild benchmarks (\Beta and \wild), which better mirror the challenges of \emph{multi-scale temporal element entanglement (i.e., multi-granular temporal coexistence and temporal interval topological disparity)} and \emph{cross-source temporal structural imbalance (i.e., multi-source temporal incompleteness and temporal event density imbalance)} in TKGs. 
Extensive experiments demonstrate the difficulty of the new benchmarks and also the effectiveness of the proposed solution, which achieves up to a 43.3\% improvement in Hits@1 over 24 strong baselines, while maintaining scalability and efficiency.

This work focuses on structural and temporal heterogeneity within and across temporal knowledge graphs. Each entity in TKGs may also contains richer, multimodal contextual signals~\cite{CVPR24EA}, such as temporal images, videos, etc. As a future direction, we plan to extend TKGA-Wild benchmarks and \ourmodel to incorporate multimodal temporal alignment tasks.

\balance

\bibliographystyle{IEEEtran}
\bibliography{ref}

\ifCLASSOPTIONcaptionsoff
  \newpage
\fi

\vfill

\end{document}